\documentclass[a4paper,11pt]{article}
\pdfoutput=1 

\usepackage{jheppub} 

\usepackage[utf8]{inputenc}

\usepackage[table]{xcolor}
\usepackage{colortbl}
\usepackage{tikz}

\usetikzlibrary{decorations.pathmorphing}
\usetikzlibrary{shapes}
\usetikzlibrary{shapes.geometric}
\usetikzlibrary{decorations.pathreplacing}
\usepackage[force]{feynmp-auto}
\usepackage{multirow,tabularx}
\usepackage{pbox}
\usepackage{slashed}

\newcommand{\pzero}{\cellcolor{green!25} 0}

\newcommand{\pnd}{\cellcolor{gray!25} $\times$}
\newcommand{\pnz}{\cellcolor{red!25} R}

\newcommand{\rotp}[1]{\rotatebox{90}{#1}}

\newcommand{\py}{\cellcolor{green!25} Y}
\newcommand{\pn}{\cellcolor{red!25} N}
\newcommand{\pyn}{\cellcolor{yellow!25} Y(n.A.)}

\newcommand{\vp}{\ensuremath{V^+}}
\newcommand{\vm}{\ensuremath{V^-}}
\newcommand{\fp}{\ensuremath{\psi^+}}
\newcommand{\fm}{\ensuremath{\psi^-}}
\newcommand{\s}{\ensuremath{\phi}}

\preprint{FERMILAB-PUB-19-646-T}

\title{Loops and trees in generic EFTs}

\author[a]{Nathaniel Craig,}
\author[b,a]{Minyuan Jiang,}
\author[c,d,e]{Ying-Ying Li,}
\author[f,a]{and Dave Sutherland}

\affiliation[a]{Department of Physics, University of California, Santa Barbara, CA 93106, USA}
\affiliation[b]{CAS Key Laboratory of Theoretical Physics, Institute of Theoretical Physics, Chinese Academy of Sciences, Beijing 100190, P.R.C}
\affiliation[c]{Theoretical Physics Department, Fermilab, Batavia, Illinois 60510, USA}
\affiliation[d]{Department of Physics, The Hong Kong University of Science and Technology, Clear Water Bay, Kowloon, Hong Kong S.A.R., P.R.C}
\affiliation[e]{Kavli Institute for Theoretical Physics, University of California, Santa Barbara, CA 93106, USA}
\affiliation[f]{INFN Sezione di Trieste, via Bonomea 265, 34136 Trieste, Italy}

\emailAdd{ncraig@physics.ucsb.edu}
\emailAdd{minyuan@itp.ac.cn}
\emailAdd{yingying@fnal.gov}
\emailAdd{dave.sutherland@sissa.it}

\abstract{We consider aspects of tree and one-loop behavior in a generic 4d EFT of massless scalars, fermions, and vectors, with a particular eye to the high-energy limit of the Standard Model EFT at operator dimensions 6 and 8. First, we classify the possible Lorentz structures of operators and the subset of these that can arise at tree-level in a weakly coupled UV completion, extending the tree/loop classification through dimension 8 using functional methods. Second, we investigate how operators contribute to tree and one-loop helicity amplitudes, exploring the impact of non-renormalization theorems through dimension 8. We further observe that many dimension 6 contributions to helicity amplitudes, including rational parts, vanish exactly at one-loop level. This suggests the impact of helicity selection rules extends beyond one loop in non-supersymmetric EFTs.

}

\newcommand{\beq}{\begin{equation*} }
\newcommand{\eeq}{\end{equation*} }
\newcommand{\dif}{\mathrm{d}}
\newcommand{\cut}{\mathrm{cut}}

\newcommand{\amp}{\mathcal{A}}
\newcommand{\lag}{\mathcal{L}}

\newcommand{\abs}[1]{| #1 |}

\newcommand{\ab}[2]{\ensuremath{\langle #1 #2 \rangle}}
\newcommand{\sqb}[2]{\ensuremath{[ #1 #2 ]}}

\begin{document} 
\maketitle
\flushbottom

\newcommand{\auv}{\ensuremath{\amp_\text{UV}}}

\section{Introduction}

The discovery of a Standard Model-like Higgs has driven considerable progress in our understanding of four-dimensional chiral effective field theories (EFTs) of scalars, fermions, and vectors, with a particular view to their relevance in describing the effective field theory of the Standard Model. Recent highlights include the ability to systematically enumerate operators in generic EFTs \cite{Henning:2015alf,Lehman:2015via}, and the revival of functional techniques \cite{Gaillard:1985uh, Chan:1986jq, Cheyette:1987qz} for matching such EFTs to their weakly coupled UV completions \cite{Henning:2014wua, Drozd:2015rsp, delAguila:2016zcb, Boggia:2016asg, Henning:2016lyp,Ellis:2016enq, Fuentes-Martin:2016uol, Zhang:2016pja, Ellis:2017jns}. Exploration of these EFTs has revealed considerable structure, imbued by physics both above them (via patterns in operator coefficients coming from the UV completion) and within them (via both mixing between EFT operators and their relation to physical observables). Characterizing this structure is vital, both as a practical matter of effectively interpreting experimental data and as a principled matter of understanding quantum field theories relevant to the real world.

Physical effects in perturbative EFTs are typically organized by a simultaneous expansion in loops and operator dimensions, with the nominal complexity growing considerably at each order. The precision attainable by the LHC and proposed future colliders provides potential access to effects arising at higher order in both expansion parameters within the Standard Model EFT, bringing tens of thousands of operators (and their ensuing radiative correlations) into play. 

Amidst this vast landscape of possibilities, attempts to develop a more comprehensive understanding of the phenomenological effects of these EFTs have led to a focus on the scattering amplitudes to which they contribute. Like renormalizable gauge theories, the lagrangian of an EFT can be somewhat cumbersome and is full of hidden redundancies, such as the ability to redefine fields without affecting S-matrix elements (which is responsible for the pernicious equation of motion relations between different lagrangian operators). The S-matrix is, by contrast, rather simple. Considering the higher dimension tree- and loop-level contributions to helicity amplitudes results in powerful `non-interference' \cite{Azatov:2016sqh} and `non-renormalization' \cite{Cheung:2015aba} theorems, with direct relevance to the structure of helicity amplitudes in the Standard Model in the high energy limit (i.e.\ above the electroweak scale). The former illuminates LHC sensitivity to new physics in diboson channels, while the latter explains the surprising pattern of zeroes \cite{Alonso:2014rga} appearing in the one-loop matrix of anomalous dimensions for dimension 6 operators in the Standard Model EFT \cite{Grojean:2013kd, Jenkins:2013zja, Jenkins:2013wua, Alonso:2013hga, Elias-Miro:2013gya, Elias-Miro:2013mua}, and more broadly illustrates the pattern of possible loop effects within the EFT itself. Progress has also been recently made in the formulation of EFT amplitudes for both massless and massive particles without reference to operators \cite{Shadmi:2018xan, Ma:2019gtx, Aoude:2019tzn, Durieux:2019siw}.

An orthogonal technique to help organize the many operators of an effective field theory is to consider the subset which can be generated at tree- and loop-level in a weakly coupled UV completion, see, e.g., \cite{Arzt:1993gz, Giudice:2007fh,Einhorn:2013kja}. Although the utility of this classification relies on perturbativity of the UV completion and must be used with care \cite{Jenkins:2013fya, Gripaios:2015qya, Gripaios:2018zrz}, it nonetheless can provide useful guidance in estimating the relative size of some physical effects and often has intriguing overlap with classification schemes based on non-interference and non-renormalization theorems. 

In both cases, the majority of progress in understanding the structure of generic EFTs has been made at the lowest nontrivial operator dimension, i.e.~dimension 6. However, the consequences of non-interference theorems, the richness of possible UV physics, and the anticipated high level of experimental precision all necessitate the further exploration of effects arising dimension 8. Some progress has been made in this direction (see e.g. \cite{Degrande:2013kka, Liu:2016idz, Liu:2019vid, Hays:2018zze, Passarino:2019yjx, Durieux:2019siw}), but the understanding of non-interference theorems, non-renormalization theorems, and the tree/loop classification up to dimension 8 is still incomplete.

 In this paper, we seek to obtain a more refined picture of the structure of generic EFTs by combining the two approaches: we study the effects of tree- and loop-level generated operators in the space of tree- and loop- level helicity amplitudes, commenting throughout on issues relevant to the Standard Model above the electroweak scale and extending results to dimension 8 wherever possible. We begin by enumerating the operators themselves; Section \ref{sec:opclass} details a simple method for enumerating the different Lorentz structures in a general EFT of scalars, fermions and vectors. This reproduces classifications arising from Hilbert series techniques \cite{Henning:2015alf,Lehman:2015via}, but has the advantage of involving the imposition of only four straightforward criteria. In Section \ref{sec:treeloop} we develop a new approach to the tree/loop classification of operator coefficients by using functional matching techniques to obtain a subset of the operators of \S\ref{sec:opclass} which may be generated at tree-level in perturbative UV completions, extending the classification to dimension 8. In Section \ref{sec:treeloopamp} we review and extend the respective works of \cite{Azatov:2016sqh} and \cite{Cheung:2015aba} on operators' tree- and loop- level contributions to helicity amplitudes. Among other things, we find that the dimension 6 parts of many one loop helicity amplitudes vanish entirely (including the rational terms), despite the existence of relevant Feynman diagrams. This is in contrast to the dimension 4 parts of one loop helicity amplitudes, whose nonzero rational parts violate all tree-level helicity selection rules, and suggests that the effects of helicity selection rules in non-supersymmetric EFTs may often extend beyond one loop.
    In Section \ref{sec:convolve} we review some aspects of the combined effects of the patterns described in \S\ref{sec:treeloop} and \S\ref{sec:treeloopamp}, to better understand the observable manifestations of weakly coupled new physics. We summarize our main conclusions in Section \ref{sec:conc}.

\section{Classification of higher dimensional operators\label{sec:opclass}}

In this manuscript we consider a gauge theory of massless scalars ($\phi$), fermions ($\psi$), and vectors ($V$). First of all, we wish to determine the possible Lorentz structures of higher dimensional operators in the lagrangian. For this purpose, it is easiest to consider the operators' constituent fields and derivatives in irreducible representations of the Lorentz group
\begin{equation}
\phi, \psi_\alpha, \bar\psi_{\dot\alpha}, F_{\alpha\beta}, \bar F_{\dot\alpha\dot\beta}, D_{\alpha\dot\alpha} \, ,
\label{eq:irrepObj}
\end{equation}
which respectively represent a scalar field, left and right handed chiral fermions, the left and right handed components of the field strength tensor, and the gauge covariant derivative. As usual, the undotted and dotted indices transform in the defining reps of $SU(2)_L \times SU(2)_R$, and $F_{\alpha\beta} \epsilon^{\alpha\beta} = \bar F_{\dot\alpha\dot\beta} \epsilon^{\dot\alpha\dot\beta} = 0$. Moreover, the objects in (\ref{eq:irrepObj}) respectively excite massless particles of definite helicity, $h$, which we write as
\begin{equation}
\phi, \psi^+, \psi^-, V^+, V^-, V^\pm ,
\end{equation}
where the respective helicities
\begin{equation}
h = 0, \frac12, -\frac12, 1, -1, \pm 1,
\end{equation}
are defined for outgoing particles. We will rely heavily on the correspondence between the fields enumerated in (\ref{eq:irrepObj}) and the helicity eigenstates they excite to classify the scattering processes to which a given operator contributes.

The forms of the dimension 4 operators in the EFT lagrangian are well known. Suppressing Lorentz indices, as well as those belonging to any internal symmetries, we write the schematic lagrangian
\begin{equation}
\lag_4 = -F^2 - \bar F^2 + i \bar \psi D \psi + (D \phi)^2 - \lambda \phi^4 - y \phi \psi \psi + \text{h.c.},
\end{equation}
where $D = \partial + i g A$, and $g$, $y$, and $\lambda$ stand for the strength of various gauge, Yukawa, and scalar quartic interactions. Furthermore, we make the simplifying assumption that there are no superrenormalizable dimension-three interactions, i.e., no scalar cubic interactions, so as to avoid introducing another mass scale which would complicate the subsequent analysis. Importantly, the Standard Model, in the unbroken phase, does not have such an interaction.

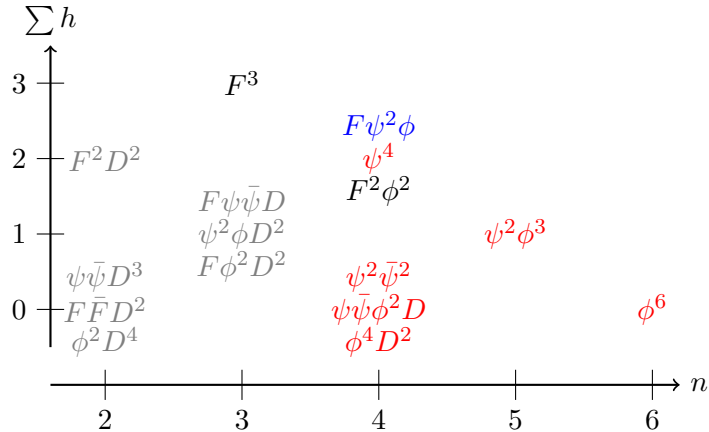
\begin{figure}
\centering
\def\eps{0.43}
\begin{tikzpicture}[x=1.8cm,y=1.0cm]
\node[gray] (d4s2) at (2,0-\eps) {$\phi^2 D^4$};
\node[gray] (d3ppb) at (2,0+\eps) {$\psi \bar \psi D^3$};
\node[gray] (d2ffb) at (2,0) {$F \bar F D^2$};
\node[red] (d2s4) at (4,0-\eps) {$\phi^4 D^2$};
\node[red]  (p2pb2) at (4,0+\eps) {$\psi^2 {\bar \psi}^2$};
\node[red]  (dppbs2) at (4,0) {$\psi \bar \psi \phi^2 D$};
\node[red]  (s6) at (6,0) {$\phi^6$};
\node[gray] (d2f2) at (2,2) {$F^2 D^2$};
\node (f2s2) at (4,2-\eps) {$F^2 \phi^2$};
\node (fp2s) at (4,2+\eps) {\textcolor{blue}{$F \psi^2 \phi$}};
\node[red]  (p4) at (4,2) {$\psi^4$};
\node[gray] (dfppb) at (3,1+\eps) {$F \psi \bar \psi D$};
\node[gray] (d2p2s) at (3,1) {$\psi^2 \phi D^2$};
\node[gray] (d2fs2) at (3,1-\eps) {$F \phi^2 D^2$};
\node (f3) at (3,3) {$F^3$};
\node[red]  (p2s3) at (5,1) {$\psi^2 \phi^3$};
\foreach \x in {2,3,4,5,6} {
\draw (\x,-0.8) -- (\x,-1.2) node [below] {\x};
}
\foreach \y in {0,1,2,3} {
\draw (1.7,\y) -- (1.5,\y) node [left] {\y};
}
\draw[->,thick] (1.6,-1) -- (6.2,-1) node [right] {$n$};
\draw[->,thick] (1.6,-0.5) -- (1.6,3.5) node [above] {$\sum h$};
\end{tikzpicture}
\caption{The classes of dimension six operators in the space of the number of fields contained, $n$, and the total helicity $\sum h$ of the particles they excite. We truncate the diagram about its axis of symmetry $\sum h = 0$. Reflect the diagram in the $n$-axis by Hermitian conjugation to obtain the full set of operators. Operators in red arise at tree level in a weakly coupled renormalizable UV completion, as defined in \S\ref{sec:treeloop}. Operators in blue and black are generated at loop level, but only those in blue are renormalized by tree-level operators. Those in gray can be expressed in terms of the others by equations of motion, when considering their effects in helicity amplitudes. \label{fig:dim6treeloop}}
\end{figure}

At higher dimensions, it is possible to form many more Lorentz invariant operators; their enumeration, taking into account the internal symmetries of the fields, is a recently solved problem \cite{Lehman:2015via,Henning:2015alf}. For the purposes of simply classifying the possible Lorentz structures of operators, it suffices to consider all products of the objects in (\ref{eq:irrepObj}) that satisfy the following three criteria:
\begin{enumerate}
\item The product contains more than one field, otherwise it is trivially a total derivative.
\item The product contains an even number of dotted and undotted indices. This is necessary to form a Lorentz invariant out of the product of fields and derivatives, because all invariant tensors of the Lorentz group ($\epsilon_{\alpha\beta}$, $\epsilon_{\dot\alpha\dot\beta}$, $\delta^\alpha_\beta$, $\delta^{\dot\alpha}_{\dot\beta}$) contract said indices in pairs.
\item If the product of fields and derivatives contains exactly two (un)dotted indices before contraction, they cannot belong to a field strength tensor $\bar F_{\dot\alpha\dot\beta}$ ($F_{\alpha\beta}$), as, after contraction, $\bar F_{\dot\alpha\dot\beta} \epsilon^{\dot\alpha\dot\beta} = F_{\alpha\beta} \epsilon^{\alpha\beta} = 0$.
\end{enumerate}
The resulting classes of operators for dimension 6 and 8 are shown in Figures \ref{fig:dim6treeloop} and \ref{fig:dim8treeloop} respectively. The operators are arranged by two coordinates,
\begin{align}
n &= \text{Number of fields,} \\
\sum h &= \text{The sum of helicities of the particles excited by the $n$ fields.}
\end{align}
Note that any vector which may be excited by a covariant derivative is ignored in the definition of these two quantities; we briefly consider such effects when constructing higher point amplitudes in Section \ref{sec:tree}. The coordinates $(n,\sum h)$ are half of the sum and difference of the holomorphic weights introduced in \cite{Cheung:2015aba}.


When considering their effect on S-matrix elements, we can entirely eliminate some classes of operators that satisfy criteria (1), (2) and (3). Dimension $d$ operators proportional to a marginal equation of motion
\begin{align}
\frac{\delta S_4}{\delta \phi} &= D^{\alpha\dot\alpha} D_{\alpha\dot\alpha} \phi + \lambda \phi^3 + y \psi \psi + y^\dagger \bar\psi \bar\psi ,\\
\frac{\delta S_4}{\delta \bar \psi^{\dot\alpha}} &= D_{\alpha\dot\alpha} \psi^\alpha - y^\dagger \phi \bar\psi_{\dot\alpha} ,\\
\frac{\delta S_4}{\delta \psi^{\alpha}} &= D_{\alpha\dot\alpha} \bar\psi^{\dot\alpha} - y \phi \psi_{\alpha} ,\\
\frac{\delta S_4}{\delta V^{\alpha\dot\alpha}} &= -D_\alpha^{\dot\beta} \bar F_{\dot\beta\dot\alpha} -D^\beta_{\dot\alpha} F_{\alpha\beta} + \bar \psi_{\dot\alpha} \psi_\alpha + \phi \overset{\leftrightarrow}{D}_{\alpha\dot\alpha} \phi ,
\end{align}
(where $S_4 = \int \lag_4$), or a Bianchi identity
\begin{equation}
0 = +D_\alpha^{\dot\beta} \bar F_{\dot\beta\dot\alpha} - D_\beta^{\dot\alpha} F_{\alpha\beta},
\end{equation}
do not contribute to $S$-matrix elements at dimension $d$ order \cite{Arzt:1993gz}. Therefore, the Lorentz structures
\begin{equation}
D^{\alpha\dot\alpha} D_{\alpha\dot\alpha} \phi ,
D_{\alpha\dot\alpha} \psi^\alpha ,
D_{\alpha\dot\alpha} \bar\psi^{\dot\alpha} ,
D_\beta^{\dot\alpha} F_{\alpha\beta} ,
D_\alpha^{\dot\beta} \bar F_{\dot\beta\dot\alpha} ,
\label{eq:zeroContactStructures}
\end{equation}
can always be eliminated in favor of products of more fields. Allowing for use of integration by parts, one may show that a term of the form (\ref{eq:zeroContactStructures}) is always present in any two or three field operator containing covariant derivatives, leading to a fourth criterion
\begin{enumerate}
\setcounter{enumi}{3}
\item If the product contains three fields or fewer, it cannot contain covariant derivatives.
\end{enumerate}
for enumerating a full set of operator classes that contribute to helicity amplitudes. Classes of operators satisfying (1), (2), and (3), but not (4), are shown in gray in Figures \ref{fig:dim6treeloop} and \ref{fig:dim8treeloop}. We will always implicitly work in a basis where such gray operators have been exchanged for lower derivative operators.

\begin{figure}
\def\eps{0.39}
\begin{tikzpicture}[x=2.2cm,y=2.5cm]
\node[gray] at (2,2) {$F^2 D^4$};
\node[gray] at (3,3) {$F^3 D^2$};
\node[gray,text width=1.6cm] at (2,0) {$F \bar F D^4$, $\phi^2 D^6$, $\psi \bar \psi D^5$};
\node[gray,text width=3.6cm] at (3,1) {$F^2 \bar F D^2$, $\phi^2 F D^4$, $\psi^2 \phi D^4$, $F \psi \bar \psi D^3$};
\node[text width=3.6cm] at (4,0) {$F^2 \bar F^2$, \textcolor{blue}{$F \bar F \phi^2 D^2$}, \textcolor{red}{$\phi^4 D^4$}, \textcolor{blue}{$\bar F \psi^2 \phi D^2$}, \textcolor{blue}{$F \bar F \psi \bar \psi D$}, \textcolor{red}{$\psi \bar \psi \phi^2 D^3$}, \textcolor{blue}{$F \bar \psi^2 \phi D^2$}, \textcolor{red}{$\psi^2 \bar \psi^2 D^2$}};
\node[text width=3.6cm] at (4,2) {$F^2 \phi^2 D^2$, \textcolor{blue}{$F \psi^2 \phi D^2$}, \textcolor{red}{$\psi^4 D^2$}, $F^2 \psi \bar \psi D$};
\node at (4,4) {$F^4$};
\node[text width=3.6cm] at (5,1) {\textcolor{red}{$F \phi^4 D^2$}, \textcolor{red}{$\psi^2 \phi^3 D^2$}, \textcolor{red}{$F\psi \bar \psi \phi^2 D$}, \textcolor{red}{$\psi^3 \bar \psi \phi D$}, \textcolor{blue}{$F^2 \bar \psi^2 \phi$}, \textcolor{red}{$F \psi^2 \bar \psi^2$}};
\node[text width=3.6cm] at (5,3) {$F^3 \phi^2$, \textcolor{blue}{$F^2 \psi^2 \phi$}, \textcolor{red}{$F \psi^4$}};
\node[text width=3.6cm] at (6,0) {\textcolor{red}{$\phi^6 D^2$}, \textcolor{red}{$\psi \bar \psi \phi^4 D$}, \textcolor{red}{$\psi^2 \bar \psi^2 \phi^2$}};
\node[text width=3.6cm] at (6,2) {\textcolor{red}{$F^2 \phi^4$}, \textcolor{red}{$F \psi^2 \phi^3$}, \textcolor{red}{$\psi^4 \phi^2$}};
\node at (7,1) {\textcolor{red}{$\psi^2 \phi^5$}};
\node at (8,0) {\textcolor{red}{$\phi^8$}};
\foreach \x in {2,3,4,5,6,7,8} {
\draw (\x,-0.8) -- (\x,-1.2) node [below] {\x};
}
\foreach \y in {0,1,2,3,4} {
\draw (1.6,\y) -- (1.4,\y) node [left] {\y};
}
\draw[->,thick] (1.5,-1) -- (8.2,-1) node [right] {$n$};
\draw[->,thick] (1.5,-0.5) -- (1.5,4.5) node [above] {$\sum h$};
\end{tikzpicture}
\caption{The classes of dimension eight operators. Operators in red arise at tree level in a weakly coupled renormalizable UV completion, as defined in \S\ref{sec:treeloop}. Operators in blue and black are generated at loop level, but only those in blue are renormalized by tree-level operators. Those in gray can be expressed in terms of the others by equations of motion, when considering their effects in helicity amplitudes. \label{fig:dim8treeloop}}
\end{figure}

Note, as an aside, that the absence of derivatives in three field operators which contribute to the S-matrix can be understood more simply from the structure of the 3-point helicity amplitude: three particle special kinematics imply either all angle bracket or all square bracket products vanish, and the amplitude is proportional to either $\ab12^a \ab23^b \ab31^c$ or $\sqb12^a \sqb23^b \sqb31^c$, for some integer $a,b,c$. It thus contains no momentum factors $p_i \sim |i\rangle [ i|$ that would arise from a partial derivative.

Using the results of \cite{Henning:2015alf}, we have checked explicitly that all of the classes of operators satisfying the above four criteria are populated at even dimensions up to 12 in the Standard Model. Having enumerated the various operator classes in a generic EFT, we now turn to characterizing the relative sizes of effects they induce, as encoded in the loop order of both the Wilson coefficients of the operators themselves and the helicity amplitudes to which they contribute.


\section{The tree vs. loop classification of operator coefficients\label{sec:treeloop}}

Any given UV completion of the generic EFT of \S\ref{sec:opclass} will generate non-trivial structure in the Wilson coefficients of the EFT operators. In weakly coupled UV completions for example, it is possible to classify the higher dimensional operators of the EFT by whether they are potentially generated at tree- or loop-level when integrating out the heavy dynamics.\footnote{Although the tree/loop classification is well-defined in weakly coupled UV completions, it may break down in strongly coupled theories and/or in the presence of super-renormalizable operators, see e.g. \cite{Jenkins:2013fya} for examples. Care must also be taken when truncating operator bases using the tree/loop classification, as operators at a given loop order in general do not form a vector subspace \cite{Gripaios:2015qya} and truncating by anticipated loop order may introduce ambiguities \cite{Gripaios:2018zrz}. Here we do not advocate for any particular application or interpretation of the tree/loop classification, but simply extend it where applicable.} The tree/loop classification of operators has been systematically explored for operators of dimension 6 in e.g.~\cite{Arzt:1994gp, Giudice:2007fh,Einhorn:2013kja}. In this section, we compactly reproduce existing work on the tree/loop classification at dimension 6 and explicitly extend the classification to dimension 8 by applying functional methods to the problem.

\newcommand{\hs}{\mathbf{\Phi}}
\newcommand{\hf}{\mathbf{\Psi}}
\newcommand{\hv}{\mathbf{V}}
\newcommand{\hh}{\mathbf{H}}
\newcommand{\rmo}{\mathrm{O}}

We consider a generic renormalizable UV theory of heavy scalars $\hs$, Weyl fermions $\hf$ and $\bar \hf$, and vectors $\hv$, all of mass $M$. Up to quadratic order in the heavy fields, it has the schematic lagrangian
\begin{align}
\lag_\text{UV} =& - \frac12 
\begin{pmatrix} \hs & \hf & \bar \hf & \hv^\mu \end{pmatrix}
\begin{pmatrix} D^2 + M^2 + \lambda \phi^2 & y \psi & y \bar \psi & 0 \\
y \psi & M + y \phi & -i \slashed{D} & 0 \\
y \bar \psi & i \slashed{D} & M + y \phi & 0 \\
0 & 0 & 0 & \pbox{38mm}{$-g_{\mu\nu}(D^2+M^2+g \phi^2) + D_\nu D_\mu - [D_\mu,D_\nu]$ }
\end{pmatrix}
\begin{pmatrix} \hs \\ \hf \\ \bar \hf \\ \hv^\nu \end{pmatrix} \nonumber \\
& -
\begin{pmatrix} \hs & \hf & \bar \hf & \hv^\mu \end{pmatrix}
\begin{pmatrix} y \psi \psi + y \bar \psi \bar \psi + \lambda \phi^3 \\ y \phi \psi \\ y \phi \bar \psi \\ g \bar \psi \sigma_\mu \psi + g \phi \overset{\leftrightarrow}{D}_\mu \phi \end{pmatrix} 
+ \rmo(\{\hs,\hf,\bar\hf,\hv \}^3) \label{eq:uvlag} \\
\equiv & -\frac12 \underline{\mathbf{H}}^T \underline{\underline{Q}} \underline{\mathbf{H}} - \underline{\mathbf{H}}^T \underline{J} + \rmo(\underline{\mathbf{H}}^3) 
\end{align}
where, again, $g$, $y$ and $\lambda$ stand for gauge, Yukawa, and scalar quartic coupling strengths. We arrange into vectors the heavy fields $\underline{\mathbf{H}}$ and their light sources $\underline{\mathbf{J}}$. We do not expand the interactions that are cubic or higher in the heavy fields, as they only contribute to operators of dimension greater than 8, whose form we do not consider in detail here. We have made two assumptions on the form of the linear and quadratic interactions of the heavy fields: one, there are no superrenormalizable interactions between the heavy and light scalars;\footnote{This simplifies the expressions in (\ref{eq:uvlag}). Including a scalar cubic would not generate any new classes of tree-level operators up to and including $d=8$.} two, there is no mixing between the heavy vector field and the scalars. Aside from these two assumptions, the form (\ref{eq:uvlag}) is completely generic in the following sense. Any other heavy-light field mixing can be diagonalized away without loss of generality (including that between heavy and light vectors \cite{delAguila:2010mx}). Note also that the form of the terms quadratic in the heavy vector can be fixed by requiring perturbative unitarity of its tree level amplitudes \cite{Henning:2014wua}.

The form of the heavy vector terms in (\ref{eq:uvlag}) can be obtained via the spontaneous breaking of a gauge boson lagrangian. Thus, our classification of renormalizable UV lagrangians aligns with that of \cite{Arzt:1994gp, Giudice:2007fh, Einhorn:2013kja}.

We perform the tree-level matching of the UV lagrangian onto that of the EFT functionally (see e.g. \cite{Henning:2014wua}). Namely, we solve the equations of motion of the heavy fields for the classical field solution
\begin{equation}
\underline{\hh}_c = - \underline{\underline{Q}}^{-1} \underline{J} + \rmo(\underline{J}^2) ,
\end{equation}
and substitute it back into the lagrangian to give the full set of tree-level operators
\begin{equation}
\lag_\text{EFT} = \frac12 \underline{J}^T \underline{\underline{Q}}^{-1} \underline{J} + \rmo(\underline{J}^3),
\end{equation}
which we expand to the desired order in $\frac1M$.

Up to dimension 6 ($\rmo(\frac1{M^2})$),
\begin{equation}
\lag_\text{EFT} = \frac1{2 M^2} \underline{J}^T 
\begin{pmatrix}
1 & 0 & 0 & 0 \\
0 & M & i \slashed{D} & 0 \\
0 & -i \slashed{D} & M & 0 \\
0 & 0 & 0 & 1
\end{pmatrix}
\underline{J} + \rmo(\frac1{M^3}) ,
\end{equation}
encoding the familiar result that all tree-level operators at dimension six are the product of two currents $J$. The tree-level classes of operators are shown in red in Figure \ref{fig:dim6treeloop}, and are exactly those which do not contain field strengths $F$/$\bar F$ (such field strengths being absent from the currents $\underline{J}$).

\newcommand{\mat}[1]{\underline{\underline{#1}}}

Working up to dimension 8, and splitting $\underline{\underline{Q}} = \underline{\underline{M}} + \underline{\underline{D}}$ into parts that contain mass factors, and covariant derivative/light field factors respectively, 
\begin{equation}
\lag_\text{EFT} = - \frac12 \underline{J}^T \mat{M}^{-1} \mat{D} \mat{M}^{-1} (1- \mat{D} \mat{M}^{-1} + \mat{D} \mat{M}^{-1} \mat{D} \mat{M}^{-1}) \underline{J} + \rmo(\frac1{M^5}) \, .
\label{eq:dim8treelevellag}
\end{equation}
Similarly to dimension 6, this expression contains all classes of dimension 8 operators that do not contain field strengths. However, some operators containing field strengths are also generated at tree-level via
\begin{enumerate}
\item the $[D_\mu,D_\nu]$ factor in $\mat{D}$ in (\ref{eq:dim8treelevellag}) (arising from the heavy vector's quadratic action) acting on a light field. This generates the operator classes $F\phi^4D^2$, $F\phi^2 \psi \bar{\psi} D$, $F\psi^2 \bar{\psi}^2$, and conjugates.
\item operators in (\ref{eq:dim8treelevellag}) which can be expressed entirely in terms of lower derivative operators via the equations of motion of the light fields, and which also contain a factor $\slashed{D} \slashed{D} \sim D^2 + \sigma^{\mu\nu} F_{\mu\nu}$ (which is equivalent to $\sim D^2 \epsilon_{\alpha\beta} + F_{\alpha\beta}$ or $D^2 \epsilon_{\dot\alpha\dot\beta} + \bar F_{\dot\alpha\dot\beta}$). This generates the operator classes $F\psi \bar {\psi} \phi^2 D$, $F\psi^2 \phi^3$, and conjugates.
\item operators in (\ref{eq:dim8treelevellag}) which cannot be expressed entirely in terms of lower derivative operators via equations of motion of the light fields, but with multiple covariant derivatives which may be anti-commuted. Here, the question is a basis dependent one: the existence of the operator with an extra field strength depends on arrangement of derivatives in the basis operator for the operator class with more derivatives (e.g. the choice of whether one uses an operator containing $\slashed{D} \slashed{D}$ or $D^2$). In the interest of generality, we therefore also include $F^2 \phi^4$ and $F \psi^4$ (and conjugates) as operators which can be generated at tree level. In any event, when it comes to considering the operators' basis independent effects in helicity amplitudes, there exist local interactions $V^+ (\psi^+)^4$, $(V^+)^2 \phi^4$ and conjugates with tree-level coefficients.
\end{enumerate}

The tree-level operator classes at dimension 8 are shown in red in Figure \ref{fig:dim8treeloop}. Whereas no operators containing $F$ are tree-level at dimension 6, at dimension 8 operators with one $F$ can be generated at tree level, other than $F\psi^4$, which depends on the choice of basis and whose effects can be accounted for by operators of the class $\psi^4 D^2$. Operators with two or more field strengths cannot be generated except for $F^2\phi^4$, which also depends on the choice of basis. All operators with no field strengths can be generated at tree level. 

Note that there is, in principle, yet another way to generate tree-level dimension 8 operators indirectly, although it does not generate any new classes of operator beyond those described above. This happens when we use equations of motion to reduce dimension 6 operators that are proportional to $D^2 \phi$, $\slashed{D} \psi$, $\slashed{D} \bar\psi$, $DF$, or $D\bar F$, into operators with fewer derivatives. Consider a field redefinition (which does not change the physics of our EFT) of dimension 6 order of a field $f$
\begin{align}
f &\rightarrow f+\frac{X}{M^2}\nonumber\\
\mathcal{L_\text{EFT}} &\rightarrow \mathcal{L_\text{EFT}}+\frac{X}{M^2}\frac{\delta S_4}{\delta f}+\frac{X^2}{M^4}\frac{\delta^2 S_4}{\delta f^2}+\frac{X}{M^2}\frac{\delta S_6}{\delta f}+ \ldots .
{\label{eq:EOM}}
\end{align}
where $S_4$ and $S_6$ denote the dimension 4 and dimension 6 part of the effective action and the ``\ldots'' denote operators of dimension higher than 8. In the second line, the first extra term reduces a dimension 6 operator 
 into operators with fewer derivatives. We must check that performing this shift does not inadvertently generate new classes of tree-level dimension 8 operators (necessarily containing field strengths) via the last two terms.

First note that for the dimension 8 operator generated in this way to be tree level, we must have that $X$ be part of a tree level dimension 6 operator, so that $\frac{X}{M^2}\frac{\delta S_4}{\delta f}$ eliminates a tree-level dimension 6 operator. We know that all tree-level dimension 6 operators do not have field strengths (and contain at least four fields). This means that --- to tree-level order --- there is no field strength in $X$, nor $\frac{\delta S_4}{\delta f}$, nor $\frac{\delta S_6}{\delta f}$. Therefore the last two terms in (\ref{eq:EOM}) contain no field strengths, and do not generate any new classes of dimension 8 operator at tree-level.

\section{The tree vs. loop classification of helicity amplitudes\label{sec:treeloopamp}}

Thus far we have considered the possible tree/loop structure of Wilson coefficients for operators in generic EFTs, generated by physics above the cutoff of the EFT and germane when the UV completion is perturbative. Equally important -- and wholly insensitive to the unknown details of UV physics -- is the tree/loop structure of helicity amplitudes within the EFT itself. 

\subsection{Tree-level helicity amplitudes\label{sec:tree}}

We begin by reviewing \cite{Azatov:2016sqh,Cheung:2015aba} the contribution of the EFT operators to tree-level dimension $d$ processes, which we define as the part of an $n$-leg helicity amplitude which scales with energy $E$ as $E^{d-n}$. A dimension $d$ process is built out of the vertices arising from operators' contact interactions. If each operator has dimension $d_i$, then
\begin{equation}
d - 4 = \sum_\text{vertices} (d_i - 4),
\label{eq:powercount}
\end{equation}
and the dimension $d$ processes thus potentially contains an arbitrary number of marginal interactions.
As per the treatment of the operators in \S\ref{sec:opclass}, it is useful to classify these processes by the number, $n$, and the total helicity, $\sum h$, of the external legs, as these coordinates combine very simply when constructing amplitudes on-shell.

Each tree-level process $C$ (that is not a simple contact interaction) will have a pole in at least one factorisation channel whose residue is proportional to the product of two smaller helicity amplitudes $A$ and $B$, as shown in Figure \ref{fig:treeRule}. The total number of external legs of $C$ is two fewer than that of $A$ and $B$ (having removed an external leg from each to form an internal line); their total helicity is the sum of the external helicities of $A$ and $B$ (having removed a leg of equal and opposite helicity from $A$ and $B$ to form an on-shell internal line). Thus, by knowing the values $n$ and $\sum h$ of all contact interactions --- which are simply related to the operators of the lagrangian --- one may recursively build a map of all possible tree-level processes in these coordinates.

\begin{figure}
\centering
\begin{tikzpicture}
\node[draw,circle,fill=gray,inner sep=8pt] (amp1) at (0,0) {$A$};
\node[draw,circle,fill=gray,inner sep=8pt] (amp2) at (3,0) {$B$};
\node[draw,circle,fill=gray,inner sep=8pt] (amp3) at (6,0) {$C$};
\node at (0,-2) {$\begin{pmatrix} n_A \\ \sum h_A \end{pmatrix}$};
\node at (1.5,-2) {$+ \!\! \begin{pmatrix} -2 \\ 0 \end{pmatrix} \!\! +$};
\node at (3,-2) {$\begin{pmatrix} n_B \\ \sum h_B \end{pmatrix}$};
\node at (6,-2) {$\begin{pmatrix} n_C \\ \sum h_C \end{pmatrix}$};
\node at (4.5,-2) {$=$};
\node at (4.5,0) {$=$};
\foreach \ang in {100,140,...,260}
  \draw[thick] (amp1) -- (\ang:1);
\foreach \ang in {-80,-40,...,80}
  \draw[thick] (amp2) -- ++(\ang:1);
\foreach \ang in {0,36,...,324}
  \draw[thick] (amp3) -- ++(\ang:1);
\draw[thick] (amp1) -- (1,0) node[label=$\pm$] {};
\draw[thick] (amp2) -- (2,0) node[label=$\mp$] {};
\draw[dashed] (1.5,-1) -- (1.5,1);
\end{tikzpicture}
\caption{The tree-level rule for constructing amplitudes in $(n,\sum h)$ space \cite{Cheung:2015aba}.\label{fig:treeRule}}
\end{figure}

Consider first the dimension 4 processes. All three-point processes, namely
\begin{equation}
\phi \psi^+ \psi^+, \phi \phi V^+, \psi^+ \psi^- V^+, V^+ V^+ V^-,
\end{equation}
and conjugates, have $\sum h = \pm 1$.\footnote{This is fixed by the dimension of the process. $\amp \sim \ab12^a \ab23^b \ab31^c$ implies $a+b+c = -\sum h = d -n=1$, whereas $\sqb12^a \sqb23^b \sqb31^c$ has $\sum h = 1$.} From 2 three-point interactions, one might naively construct the following processes at $(n,\sum h) = (4,2)$:
\begin{equation}
\phi \psi^+ \psi^+ V^+,
\psi^+ \psi^- V^+ V^+,
\phi \phi V^+ V^+,
V^+ V^+ V^+ V^- \, .
\end{equation}
However, these are identically zero on shell \cite{Cheung:2015aba}. In fact, the only non-zero process at $(4,2)$ is the all-positive fermion amplitude $(\psi^+)^4$. This holds, furthermore, when accounting for the two four-point contact interactions at dimension 4: the scalar quartic, and the Yang-Mills four vector vertex both contribute exclusively to $(4,0)$ at four-point (the latter is also simply zero in an appropriate choice of gauge, and otherwise accounted for by the process comprising two 3-point Yang-Mills vertices). Similarly the only dimension 4 tree-level amplitude at $(4,-2)$ is $(\psi^-)^4$.

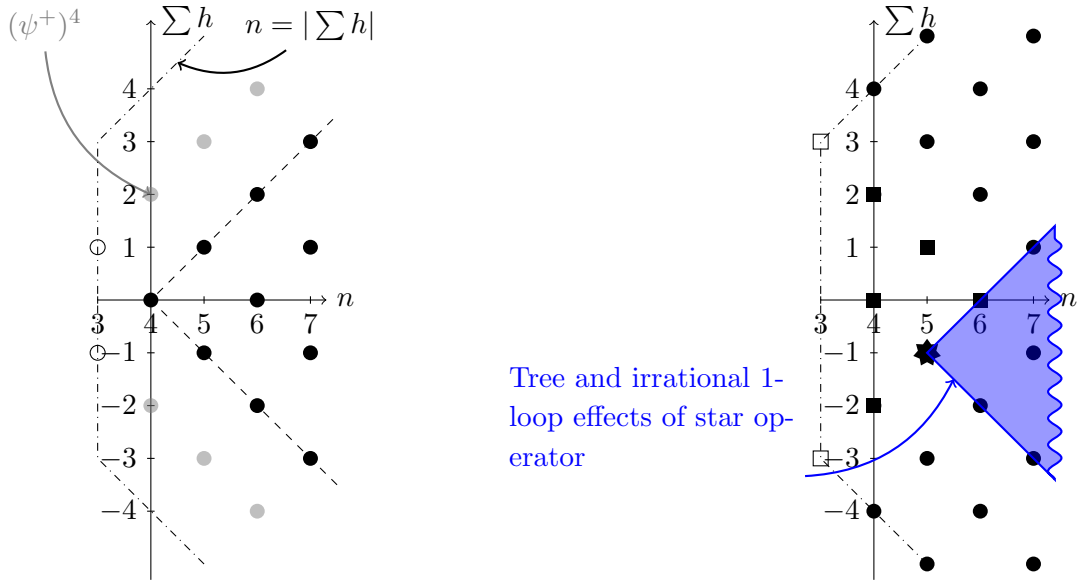
\begin{figure}
\begin{tikzpicture}[scale=0.7]
\draw[->] (3,0) -- (7.3,0) node[right] {$n$};
\draw[->] (4,-5.3) -- (4,+5.3) node[right] {$\sum h$};
\node[gray] (psi) at (2,5.2) {$(\psi^+)^4$};
\path[->,thick,gray] (psi) edge [bend right] (4,2);
\node (ext) at (7,5.2) {$n = |\sum h|$};
\path[->,thick] (ext) edge [bend left] (4.5,4.5);
\foreach \x/\xtext in {3/3, 4/4, 5/5, 6/6, 7/7}
    \draw (\x,2pt) -- (\x,-2pt) node[below] {$\xtext$};
\foreach \y/\ytext in {-4/-4,-3/-3, -2/-2, -1/-1, 1/1,2/2,3/3,4/4}
    \draw[shift={(4,0)}] (2pt,\y) -- (-2pt,\y) node[left] {$\ytext$};
\foreach \x/\y in {4/0,5/-1,5/1,6/-2,6/0,6/+2,7/-3,7/-1,7/1,7/3}
    \fill[black] (\x,\y) circle (4pt);
\foreach \x/\y in {3/1,3/-1}
    \draw[black] (\x,\y) circle (4pt);
\foreach \x/\y in {4/2,4/-2,5/3,5/-3,6/4,6/-4}
    \fill[gray,opacity=0.5] (\x,\y) circle (4pt);
\draw[dashed] (7.5,-3.5) -- (4,0) -- (7.5,3.5);
\draw[dash dot] (5,-5) -- (3,-3) -- (3,3) -- (5,5);
\end{tikzpicture}\hspace{15mm}\begin{tikzpicture}[scale=0.7]
\draw[->] (3,0) -- (7.3,0) node[right] {$n$};
\draw[->] (4,-5.3) -- (4,+5.3) node[right] {$\sum h$};
\foreach \x/\xtext in {3/3, 4/4, 5/5, 6/6, 7/7}
    \draw (\x,2pt) -- (\x,-2pt) node[below] {$\xtext$};
\foreach \y/\ytext in {-4/-4,-3/-3, -2/-2, -1/-1, 1/1,2/2,3/3,4/4}
    \draw[shift={(4,0)}] (2pt,\y) -- (-2pt,\y) node[left] {$\ytext$};
\foreach \x/\y in {4/-2,4/0,4/2,5/-1,5/1,6/0}
    \fill[black,xshift=-4pt,yshift=-4pt] (\x,\y) rectangle ++(8pt,8pt);
\foreach \x/\y in {3/3,3/-3}
    \draw[black,xshift=-4pt,yshift=-4pt] (\x,\y) rectangle ++(8pt,8pt);
\foreach \x/\y in {4/2,4/4,4/-2,4/-4,5/5,5/3,5/-3,5/-5,6/4,6/2,6/-4,6/-2,7/-5,7/-3,7/-1,7/1,7/3,7/5}
    \fill[black] (\x,\y) circle (4pt);
\draw[dash dot] (5,-5) -- (3,-3) -- (3,3) -- (5,5);
\node[fill=black,star,star points=7,inner sep=2.5pt] at (5,-1) {};
\draw[thick,blue,fill=blue,fill opacity=0.4] decorate[decoration=snake]{(7.4,-3.4) -- (7.4,1.4)} -- (5,-1) -- cycle;
\node[blue,text width=4cm] (ren) at (0,-2.2) {Tree and irrational 1-loop effects of star operator};
\path[->,thick,blue] (ren) edge [bend right] (5.5,-1.5);
\end{tikzpicture}
\caption{Dimension 4 (left) and dimension 6 (right) amplitudes. Squares denote possible contact insertions of operators. The starred contact insertion at $(5,-1)$ may only contribute to tree-level and cut constructible one-loop level (barring exceptional amplitudes) processes within the blue cone.\label{fig:dim4map}}
\end{figure}

Due to the absence of contact interactions at 5-point and above, all $n \geq 5$ dimension 4 tree-level processes may be constructed recursively from those at 3- and 4-point, using the tree-level rule of Figure \ref{fig:treeRule}; we mark their possible coordinates with circles in the left panel of Figure \ref{fig:dim4map}. Note that the majority of four or higher point tree-level processes lie on or within the dashed cone $\abs{\sum h} = n-4$, whose boundary demarcates, amongst others, the location of the all-gluon maximally helicity violating amplitudes. The processes outside the cone, shown in grey in Figure \ref{fig:dim4map}, necessarily factorise into the 4-point all-plus or all-minus fermion amplitude, and we term these `exceptional amplitudes'. Note that such processes in the Standard Model are suppressed by the product of up- and down-type Yukawas.

The observation that nearly all of the dimension 4 amplitudes lie on or within the dashed `MHV cone' of Figure \ref{fig:dim4map} drives many of the following statements in this paper regarding the tree- and one-loop structures in the space of $(n,\sum h)$.

For higher dimensional processes, by contrast, we are unaware of any such helicity selection rules causing constructed tree-level amplitudes to vanish on-shell. Consider the contribution of a given dimension $d>4$ operator to dimension $d$ processes. By power counting (\ref{eq:powercount}), they comprise a contact insertion of the dimension $d$ operator in question, plus an optional arbitrary number of dimension 4 vertices. The process which is simply a contact insertion of the operator has the same $(n,\sum h)$ coordinate as the operator, as defined in Section \ref{sec:opclass}. Adding dimension 4 interactions, by the tree level rule of Figure \ref{fig:treeRule}, leads to a succession of processes with larger $n$ that lie within a cone, whose apex is at the original $(n,\sum h)$ coordinate of the operator. We depict such a cone in blue for the starred operator in Figure \ref{fig:dim4map}. This cone contains all tree-level dimension $d$ amplitudes proportional to the Wilson coefficient of the operator in question \cite{Azatov:2016sqh}. We note here that the helicity amplitude generated when a covariant derivative in a local operator excites a vector is the same as that generated when the external leg on which the covariant derivative acts is dressed with a dimension 4 gauge vertex, and these effects are therefore included within the cone.

Figure \ref{fig:dim4map} represents a map of the possible tree-level helicity amplitudes of dimension 6 processes. Square markers show the locations of contact insertions of the operators of Figure \ref{fig:dim6treeloop}. Note how, in the absence of helicity selection rules, the processes fill the space up to the line $\abs{\sum h} = n$, which marks the location of the all plus and all minus vector amplitudes (which, at dimension 6, are necessarily proportional to the $F^3$ coefficient).


The situation is identical in higher dimensions. The map of dimension $d > 6$ (for even $d$) amplitudes is the same, other than that there are no three-point amplitudes, and there are more square markers denoting contact interactions (which lie in a region bounded by the lines $n=4$, $\abs{\sum h} = n$, and $\abs{\sum h} = d-n$). 

With the tree-level maps in hand, we turn to construct the one-loop amplitudes in the general (massless) EFT.

\subsection{One-loop helicity amplitudes\label{sec:loop}}

Consider the one-loop contributions of a given dimension $d$ operator to dimension $d$ processes. We reprise the argument in \cite{Cheung:2015aba} after reviewing some aspects of generalised unitarity methods. Applied explicitly to the operator classification of \S\ref{sec:opclass}, this leads us to extend the results of \cite{Cheung:2015aba} to operators of dimension 8.

Any one-loop amplitude can be written in the form \cite{Ossola:2006us}
\begin{equation}
O^\text{1-loop} = \int \frac{\dif^D l}{(2\pi)^D} \amp^\text{1-loop} = \int \frac{\dif^D l}{(2\pi)^D} \frac{\mathcal{N}(l)}{\bar D_1 \bar D_2 \ldots \bar D_N}
\label{eq:1loopamp}
\end{equation}
where the numerator $\mathcal{N}(l)$ is a polynomial function of the loop momentum, $l$, and external momenta and polarisations. The scalar propagator denominators $\bar D_i = (l+q_i)^2$, for some linear combinations of external particle momenta $q_i$.

The four-dimensional part of the numerator $\mathcal{N}(l)$ (obtained when its constituent $d$-dimensional objects are replaced by their four dimensional components) can be parametrised by \cite{Ossola:2006us}
\begin{align}
\mathcal{N}(l) |_\text{4d} =& \sum_{i_1} (a(i_1) + \tilde a(i_1;l)) \prod_{j \neq i_1} D_j \nonumber \\
&+ \sum_{i_1,i_2} (b(i_1,i_2) + \tilde b(i_1,i_2;l)) \prod_{j \neq i_1,i_2} D_j \nonumber \\
&+ \sum_{i_1,i_2,i_3} (c(i_1,i_2,i_3) + \tilde c(i_1,i_2,i_3;l)) \prod_{j \neq i_1,i_2,i_3} D_j \nonumber \\
&+ \sum_{i_1,i_2,i_3,i_4} (d(i_1,i_2,i_3,i_4) + \tilde d(i_1,i_2,i_3,i_4;l)) \prod_{j \neq i_1,i_2,i_3,i_4} D_j .
\label{eq:numDecomp}
\end{align}
In (\ref{eq:numDecomp}), the sums run over distinct subsets, of the appropriate size, of the integers $1$ to $N$. $D_i$ are the denomiators $\bar D_i$ when restricted to four dimensions. The Passarino-Veltman coefficients $a$, $b$, $c$, and $d$ are rational functions of the external momenta and polarisations; their counterparts with tildes --- the spurious terms --- are rational functions of the external momenta, polarisations, and the loop momentum $l$, which vanish upon integration with respect to the measure $\int \dif^4 l$. $\tilde d$, for instance, is necessarily a single term of the form
\begin{equation}
\tilde d(i_1,i_2,i_3,i_4;l) = \tilde d(i_1,i_2,i_3,i_4) (l \cdot n_4)
\end{equation}
where $\tilde d(i_1,i_2,i_3,i_4)$ is independent of the loop momentum, and $n_4$ is a four vector orthogonal to $q_{i_2}-q_{i_1}$, $q_{i_3}-q_{i_1}$, and $q_{i_4}-q_{i_1}$. The number of $l$-dependent structures in $\tilde a$, $\tilde b$, and $\tilde c$ that are necessary to parametrise the $l$ dependence of $\mathcal{N}(l)$ is, however, a function of the order of $\mathcal{N}(l)$ \cite{Ellis:2011cr}.

Performing the integration in $D=4-2\epsilon$ dimensions, we obtain the Passarino-Veltman decomposition \cite{Passarino:1978jh}
\begin{align}
O^\text{1-loop} = \sum_{i_1} a(i_1) I_1(i_1) + \sum_{i_1,i_2} b(i_1,i_2) I_2(i_1,i_2) + \sum_{i_1,i_2,i_3} c(i_1,i_2,i_3) I_3(i_1,i_2,i_3)  \nonumber \\
+ \sum_{i_1,i_2,i_3,i_4} d(i_1,i_2,i_3,i_4) I_4(i_1,i_2,i_3,i_4) + R ,
\label{eq:PVDecomp}
\end{align}
written in terms of the scalar integrals
\begin{equation}
I_k(i_1,\ldots,i_k) = \int \frac{\dif^d l}{(2\pi)^d} \frac{1}{\bar D_{i_1} \ldots \bar D_{i_k} } .
\end{equation}
The rational part $R$ of $O^\text{1-loop}$ is a rational function of the external momenta and polarisations. It arises in the above decomposition from integration over the $D-4$ dimensional parts of the numerator $\mathcal{N}$, as well as over ratios of $\frac{D_i}{\bar D_i}$ resulting from the substitution (\ref{eq:numDecomp}) \cite{Ossola:2008xq}. The rational parts considered in \S\ref{non-interf} are of the former type.

One defines $k=2,3,4$ particle cuts of the integrand of (\ref{eq:1loopamp}) by restricting it to the locus of (complex) 4d loop momenta $l$ for which $k$ internal propagators are on-shell:
\begin{equation}
\cut_{i_1 \ldots i_k} \amp^\text{1-loop} = \amp^\text{1-loop} D_{i_1} \ldots D_{i_k} \Big|_{D_{i_1} = \ldots = D_{i_k} = 0} .
\end{equation}
When applied to the decomposition (\ref{eq:numDecomp}), we obtain
\begin{align}
\cut_{i_1 i_2} \amp^\text{1-loop} =& (b(i_1,i_2) + \tilde b(i_1,i_2;l)) + \sum_{i_3} (c(i_1,i_2,i_3) + \tilde c(i_1,i_2,i_3;l)) \frac{1}{D_{i_3}} \nonumber \\
&+ \sum_{i_3,i_4} (d(i_1,i_2,i_3,i_4) + \tilde d(i_1,i_2,i_3,i_4;l)) \frac{1}{D_{i_3} D_{i_4}} , \label{eq:2cutPV} \\
\cut_{i_1 i_2 i_3} \amp^\text{1-loop} =& (c(i_1,i_2,i_3) + \tilde c(i_1,i_2,i_3;l)) + \sum_{i_4} (d(i_1,i_2,i_3,i_4) + \tilde d(i_1,i_2,i_3,i_4;l)) \frac{1}{D_{i_4}} , \label{eq:3cutPV} \\
\cut_{i_1 i_2 i_3 i_4} \amp^\text{1-loop} =& (d(i_1,i_2,i_3,i_4) + \tilde d(i_1,i_2,i_3,i_4;l)) .\label{eq:4cutPV} 
\end{align}
If one is able to evaluate the left hand sides of (\ref{eq:2cutPV}--\ref{eq:4cutPV}) at sufficiently many different values of $l$, one may then uniquely solve for the coefficients $b$, $c$, $d$, $\tilde b$, $\tilde c$, and $\tilde d$ by linear algebra.

Consider now the two particle cut of a dimension $d$ process, which contains a single insertion of a dimension $d$ operator. The integrand, under the conditions that internal legs $i_1$ and $i_2$ be on-shell, factorises into a product of tree level processes --- one of them dimension $d$, one of them dimension $4$:
\begin{equation}
\cut_{i_1 i_2} \amp^\text{1 loop} = \amp^\text{tree, dim $d$} \amp^\text{tree, dim $4$} .
\end{equation}
Moreover, the $(n,\sum h)$ coordinates of the one-loop process are fixed by those of the factorised tree-level pieces, as per Figure \ref{fig:looprule}.

Assume that both $\amp^\text{tree, dim $d$}$ and $\amp^\text{tree, dim $4$}$ are 4-point or greater, i.e. $(q_{i_1}-q_{i_2})^2 \not\equiv 0$. Unless it is an exceptional amplitude, $\amp^\text{tree, dim $4$}$ lives within the `MHV cone' $\abs{\sum h} \leq n-4$ whose apex is $(4,0)$. By the loop level construction rule of Figure \ref{fig:looprule}, this implies $\cut_{i_1 i_2} \amp^\text{1 loop}$ is within the cone whose apex is the coordinates of $\amp^\text{tree,dim $d$}$. By the tree level considerations of Section \ref{sec:tree}, $\amp^\text{tree,dim $d$}$, and by transitivity $\cut_{i_1 i_2} \amp^\text{1 loop}$, is within the cone of the contact insertion of the dimension $d$ operator, see again the blue cone in Figure \ref{fig:dim4map}.

\begin{figure}
\centering
\begin{tikzpicture}
\node[draw,circle,fill=gray,inner sep=8pt] (amp1) at (0,0) {$A$};
\node[draw,circle,fill=gray,inner sep=8pt] (amp2) at (3,0) {$B$};
\node[draw,circle,fill=gray,inner sep=8pt] (amp3) at (6,0) {$C$};
\node at (0,-2) {$\begin{pmatrix} n_A \\ \sum h_A \end{pmatrix}$};
\node at (1.5,-2) {$+ \!\! \begin{pmatrix} -4 \\ 0 \end{pmatrix} \!\! +$};
\node at (3,-2) {$\begin{pmatrix} n_B \\ \sum h_B \end{pmatrix}$};
\node at (6,-2) {$\begin{pmatrix} n_C \\ \sum h_C \end{pmatrix}$};
\node at (4.5,-2) {$=$};
\node at (4.5,0) {$=$};
\foreach \ang in {100,140,...,260}
  \draw[thick] (amp1) -- (\ang:1);
\foreach \ang in {-80,-40,...,80}
  \draw[thick] (amp2) -- ++(\ang:1);
\foreach \ang in {0,36,...,324}
  \draw[thick] (amp3) -- ++(\ang:1);
\draw[thick] (amp1) to [out=45,in=180] (1,0.8) node[label=$\pm$] {};
\draw[thick] (amp1) to [out=-45,in=180] (1,-0.8) node[label=$\pm$] {};
\draw[thick] (amp2) to [out=135,in=0] (2,0.8) node[label=$\mp$] {};
\draw[thick] (amp2) to [out=-135,in=0] (2,-0.8) node[label=$\mp$] {};
\draw[dashed] (1.5,-1) -- (1.5,1);
\end{tikzpicture}
\caption{The one-loop level rule for constructing amplitudes in $(n, \sum h)$ space \cite{Cheung:2015aba}. \label{fig:looprule}}
\end{figure}

Consider the opposite case: a $\amp^\text{1 loop}$ which is outside the cone of the dimension $d$ operator present within the amplitude. In the absence of exceptional amplitudes in its factorisation channels
\begin{equation}
\cut_{i_1 i_2} \amp^\text{1 loop} = 0, \quad \forall i_1, i_2 \text{ where } (q_{i_1}-q_{i_2})^2 \not\equiv 0 .
\label{eq:noTwoPartMassiveCut}
\end{equation}
All three and four particle cuts may be written without loss of generality as
\begin{align}
\cut_{i_1 i_2 i_3} \amp^\text{1-loop} = (\cut_{i_1 i_2} \amp^\text{1-loop}) D_{i_3} \Big|_{D_{i_1} = D_{i_2} = D_{i_3} = 0} \\
\cut_{i_1 i_2 i_3 i_4} \amp^\text{1-loop} = (\cut_{i_1 i_2} \amp^\text{1-loop}) D_{i_3} D_{i_4} \Big|_{D_{i_1} = D_{i_2} = D_{i_3} = D_{i_4} = 0}
\end{align}
where we choose $i_1$ and $i_2$ such that $\abs{i_1 - i_2} \mod N > 1$ and thus $(q_{i_1}-q_{i_2})^2 \not\equiv 0$ (i.e.\ we use a two particle cut of non-adjacent internal legs). In turn, this implies
\begin{align}
\cut_{i_1 i_2 i_3} \amp^\text{1 loop} = 0, \quad \forall i_1, i_2, i_3, \\
\cut_{i_1 i_2 i_3 i_4} \amp^\text{1 loop} = 0, \quad \forall i_1, i_2, i_3, i_4 .
\end{align}
By the now homogeneous equations (\ref{eq:2cutPV}--\ref{eq:4cutPV}), we conclude
\begin{align}
b(i_1,i_2) = \tilde b(i_1,i_2;l) = 0 \quad \forall i_1, i_2, (q_{i_1}-q_{i_2})^2 \not\equiv 0  \\
c(i_1,i_2,i_3) = \tilde c(i_1,i_2,i_3;l) = 0 \quad \forall i_1, i_2, i_3, \\
d(i_1,i_2,i_3,i_4) = \tilde d(i_1,i_2,i_3,i_4;l) = 0 \quad \forall i_1, i_2, i_3, i_4.
\end{align}
The integrated amplitude is then
\begin{align*}
O^\text{1-loop} &= \sum_{i_1} a(i_1) I_1(i_1) + \sum_{i_1,i_2,(q_{i_1}-q_{i_2})^2 \equiv 0} b(i_1,i_2) I_2(i_1,i_2) + R \\
&= R,
\end{align*}
the last equality following because, in dimensional regularisation, scaleless tadpoles and bubbles vanish.\footnote{Even in other regularisation schemes, where the scaleless integrals are non-zero, they are still rational in the external momenta.}

As was shown by this method in \cite{Cheung:2015aba}, a dimension $d$ operator's one-loop contributions to processes outside its cone --- exceptional cases aside --- are UV finite. Thus, as a rule, a dimension $d$ operator only renormalises other dimension $d$ operators within its cone (i.e. it renormalises those operators that can provide a contact insertion of a counterterm to absorb the divergence). A glance at the operator distributions in Figure \ref{fig:dim8treeloop}, for example, shows many dimension 8 operators are out of reach of each other, and that the well-known sparsity of the anomalous dimension matrix at dimension 6 persists at dimension 8, as it should too for even higher dimensions.

A simple corollary of \cite{Cheung:2015aba} is that an operator's one-loop contributions outside its cone --- exceptional cases aside --- are not just UV finite, but entirely rational. In short this is because the dimension 4 tree-level amplitude, obtained by cutting the one-loop amplitude in any scattering channel, always vanishes due to the aforementioned helicity selection rules. By the Cutkosky rules, this means that all discontinuities in the one-loop amplitude's kinematic invariants must vanish, which suffices to require all irrational parts to vanish \cite{Bern:1994zx}.

The absence of irrational parts has significant phenomenological implications. The absence of UV divergences means there are no logarithmic enhancement effects of the form $\log (\frac{M}{E})$, due to running down from some UV scale $M$ to that of the scattering process in the EFT, $E$. The lack of IR divergences similarly rules out enhancements of the form $\log(\frac{E}{m})$, $m$ being some small mass scale or IR cutoff. Further irrationalites in purely finite processes, proportional to the logarithm of some ratio of kinematic invariants, could have also enhanced the process in particular corners of phase space.

Thus, all logarithmic enhancements proportional to a given higher dimensional operator's coefficient are absent in many helicity amplitudes where one would expect them, diagrammatically, to be present. In the following subsection, we examine how the rational parts of the one-loop amplitude often also confound expectations, and vanish.

\subsection{Absent rational terms at dimension 6\label{non-interf}}

\begin{table}
\centering
\begin{tabular}{c | c c}
$(4,0)$ process & d.\ 4 & d.\ 6 \\
\vp \vp \vm \vm & \pyn & \pn \\
\vp \vm \fp \fm & \py & \pn \\
\vp \vm \s \s & \py & \pn \\
\vp \fm \fm \s & \py & \pn \\
\fp \fp \fm \fm & \py & \py \\
\fp \fm \s \s & \py & \py \\
\s \s \s \s & \py & \py 
\end{tabular}\hspace*{3ex}
\begin{tabular}{c | c c}
$(4,2)$ process & d.\ 4 & d.\ 6 \\
\vp \vp \vp \vm & \pn & \pyn \\
\vp \vp \fp \fm & \pn & \pyn \\
\vp \vp \s \s & \pn & \py \\
\vp \fp \fp \s & \pn & \py \\
\fp \fp \fp \fp & \py & \py 
\end{tabular}
\caption{A list of the $(n,h) = (4,0)$ and $(4,2)$ processes, whether they contain dimension 4 (d.\ 4) and dimension 6 (d.\ 6) pieces, and whether this occurs exclusively for non-Abelian vectors (n.A.).\label{tab:4pointtree}}
\end{table}

Of the possible four-point helicity amplitudes, few receive both a dimension 4 and dimension 6 contribution at tree level. The possibilities are enumerated in Table \ref{tab:4pointtree}. In short, 4-point processes with at least one external vector have contributions of only one dimension: $(4,0)$ processes with vectors have only a dimension 4 tree-level piece; $(4,\pm 2)$ processes only a dimension 6 piece. This means that there is no interference term between the two when calculating event rates \cite{Azatov:2016sqh}.

The processes in Table \ref{tab:4pointtree} that are absent at tree level are also purely rational at one loop. To see this quickly, consider the most general Passarino-Veltman decomposition of a massless 4-point one-loop amplitude in dimension regularisation (i.e.\ ignoring tadpoles and scaleless bubbles),
\begin{equation}
O^\text{1-loop} = \sum_{x \in \{s,t,u\} } \left( b_x I_2(x) + c_x I_3^{1m}(x) \right) + d I_4^{0m}(s,t) + R \, .
\end{equation}
The UV divergences must vanish, as there is no tree-level process involving a counterterm to absorb them. Thus the coefficients of the UV divergent bubbles $I_2(x)$ must be zero. The IR divergences must vanish, as there is no tree-level contribution which could be dressed with soft or collinear particles to absorb them (in cross sections) \cite{Dixon:1993xd,Azatov:2016sqh}. Thus the coefficients of the one-mass triangles $I_3^{1m}(x)$ and the zero mass box $I_4(s,t)$ also vanish, leaving a purely rational piece.

The rational parts of one-loop processes are relatively ubiquitous. As we have described, they are typically present in any one-loop amplitude, regardless of whether helicity selection rules forbid the existence of irrational parts. For example, all 4-point dimension 4 processes absent at tree-level generically have a non-zero dimension 4 rational contribution at one-loop.\footnote{One loop dimension 4 contributions to $V^+V^+V^+V^+, V^+ V^+ V^+ V^-, V^+ V^+ \psi^+ \psi^-, V^+ V^+ \phi \phi$ were computed in e.g. \cite{Bern:1991aq, Kunszt:1993sd, Dixon:1998py}. Although the one loop dimension 5 rational contributions to $V^+ \psi^+ \psi^+ \phi$ are computed in \cite{Berger:2006sh}, we are unaware of a corresponding result in the literature for the one loop dimension 4 part, but we have verified that the rational part for this amplitude is nonzero as well.} Consequently, the effects of tree-level helicity selection rules in the Standard Model extend no further than one loop, where they influence the matrix of one-loop anomalous dimensions. On these grounds, one might expect that all dimension 6 processes that are forbidden at tree level also possess nonzero rational parts at one loop. However, we find by explicit calculation that this is not the case. A substantial number of dimension 6 processes that are forbidden at tree level are also entirely zero at one loop despite the naive existence of relevant Feynman diagrams. This suggests that the impact of helicity selection rules extends beyond one loop for dimension 6 processes.

Whether the rational parts of said one-loop, dimension-6 helicity amplitudes are nonzero is of particular relevance to the radiative fate of non-interference theorems involving dimension 4 and dimension 6 tree level amplitudes, as mentioned above \cite{Azatov:2016sqh}. Insofar as the $(4,2)$ dimension 4 amplitudes with external vectors have a nonzero rational part at one loop, the non-interference theorems can be violated by the interference between one-loop dimension 4 amplitudes and tree level dimension 6 amplitudes. However, it is worth inquiring whether the same can be said for the interference between one-loop dimension 6 amplitudes and tree level dimension 4 amplitudes. These could be the leading source of radiative violations of the non-interference theorems if the dimension-6 Wilson coefficients are sizeable.

To this end, we calculate explicitly the rational contributions of $(4,0)$ dimension 6 operators to $(4,0)$ processes with external vectors, namely,
\begin{equation}
V^+ V^+ V^- V^-, V^+ V^- \phi \phi, V^+ V^- \psi^+ \psi^-, V^+ \psi^- \psi^- \phi,
\label{4noninterf}
\end{equation}
as well as
\begin{equation}
V^+ V^+ V^+ V^-, V^+ V^+ \psi^+ \psi^-,
\label{42noninterf}
\end{equation}
which also lack a tree-level dimension 6 part in a purely Abelian theory, see Table \ref{tab:4pointtree}. We do so in four simplified models described in Appendix \ref{app:simpmodels} --- two Abelian and two non-Abelian --- with the help of \texttt{FORM} \cite{Kuipers:2012rf}, \texttt{FeynArts} \cite{Hahn:2000kx}, and \texttt{FormCalc} \cite{Hahn:1998yk}, employing \texttt{FormCalc}'s default prescription of a $\gamma_5$ that anticommutes with $D$-dimensional gamma matrices (our results should, however, be scheme independent). We summarise the results here in condensed Abelian and non-Abelian tables \ref{tab:rationalResults}, where `$\times$' denotes no diagram\footnote{This means specifically no diagram for the process before the helicities of the external particles are fixed.} in either model of the type considered, `0' denotes zero rational parts in models with diagrams, and `R' denotes a non-zero rational part in at least one of the models.

\begin{table}[t]
\caption{One-loop contributions of various dimension 6 operators to certain helicity amplitudes. The upper left and right quadrants correspond to contributions of tree-level dimension 6 operators to $|h| = 0$ and select $|h| = 2$ amplitudes, respectively; the lower left and right quadrants correspond to contributions of loop-level or exceptional dimension 6 operators to $|h| = 0$ and select $|h| =2$ amplitudes. Here \textcolor{gray}{$\times$} means there's no diagram, \textcolor{green}{0} a vanishing contribution, and \textcolor{red}{R} a non-vanishing rational contribution.\label{tab:rationalResults}}

\vspace*{2ex}

\hspace*{-3ex}
\begin{minipage}{.57\textwidth}
\centering
\textbf{Non-Abelian}

\vspace*{1ex}

\begin{tabular}{c c c c c c | c c}
& & \multicolumn{4}{c|}{$(4,0)$} & \multicolumn{2}{c}{$(4,2)$}  \\
& & \rotp{$V^+ V^+ V^- V^-$} 
 & \rotp{$V^+ V^- \psi^+ \psi^-$}
 & \rotp{$V^+ V^- \phi \phi$}
 & \rotp{$V^+ \psi^- \psi^- \phi$}
 & \rotp{$V^+ V^+ V^+ V^-$}
 & \rotp{$V^+ V^+ \psi^+ \psi^- $}
 \\
\multirow{3}*{$(4,0)$} & $\psi^2 \bar \psi^2$ & \pnd & \pzero & \pnd & \pzero ${ }^*$
 & \pnd & \pnz \\
& $\phi^4 D^2$ & \pnd & \pnd & \pzero & \pnd 
 & \pnd & \pnd \\
& $\phi^2 \psi \bar \psi D$ & \pnd & \pzero & \pzero & \pzero 
 & \pnd & \pnz \\
 \hline
\multirow{3}*{$(4,2)$} & $F \psi^2 \phi$ & \pnd & \pnz & \pnz & \pnz
 & \pnd & \pzero \\
& $F^2 \phi^2$ & \pnz & \pzero & \pnz & \pnz
 & \pzero ${ }^*$& \pzero ${ }^*$\\
& $\psi^4$ & \pnd & \pzero & \pnd & \pzero
 & \pnd & \pzero \\
 \hline
\multirow{3}*{$(4,-2)$} & $\bar F \bar \psi^2 \phi$ & \pnd & \pnz & \pnz & \pnz
 & \pnd & \pzero \\
& $\bar F^2 \phi^2$ & \pnz & \pzero & \pnz & \pnz
 & \pzero & \pzero \\
& $\bar \psi^4$ & \pnd & \pzero & \pnd & \pnz
 & \pnd & \pzero \\
\end{tabular}\end{minipage}\begin{minipage}{.4\textwidth}
\centering
\textbf{Abelian}

\vspace*{1ex}

\begin{tabular}{c c c c c | c c}
 & \multicolumn{4}{c|}{$(4,0)$} & \multicolumn{2}{c}{$(4,2)$}  \\
 & \rotp{$V^+ V^+ V^- V^-$} 
 & \rotp{$V^+ V^- \psi^+ \psi^-$}
 & \rotp{$V^+ V^- \phi \phi$}
 & \rotp{$V^+ \psi^- \psi^- \phi$}
 & \rotp{$V^+ V^+ V^+ V^-$}
 & \rotp{$V^+ V^+ \psi^+ \psi^- $}
 \\
 $\psi^2 \bar \psi^2$ & \pnd & \pzero & \pnd & \pzero ${ }^*$
 & \pnd & \pzero \\
 $\phi^4 D^2$ & \pnd & \pnd & \pzero & \pnd 
 & \pnd & \pnd \\
 $\phi^2 \psi \bar \psi D$ & \pnd & \pzero & \pzero & \pzero 
 & \pnd & \pzero \\
 \hline
 $F \psi^2 \phi$ & \pnd & \pnz & \pnz & \pnz
 & \pnd & \pzero \\
 $F^2 \phi^2$ & \pnz & \pzero & \pnz & \pnz
 & \pzero & \pzero \\
 $\psi^4$ & \pnd & \pzero & \pnd & \pzero
 & \pnd & \pzero \\
 \hline
 $\bar F \bar \psi^2 \phi$ & \pnd & \pnz & \pnz & \pnz
 & \pnd & \pzero \\
 $\bar F^2 \phi^2$ & \pnz & \pzero & \pnz & \pnz 
 & \pzero & \pzero \\
 $\bar \psi^4$ & \pnd & \pzero & \pnd & \pnz
 & \pnd & \pzero \\
\end{tabular}
\end{minipage}

\end{table}%

Most of the many zeroes in Table \ref{tab:rationalResults} can be understood from the possible Lorentz structures of the rational parts. The rational part cannot have a trivial denominator, as this would imply the existence of a dimension 6 contact term that could contribute to the process considered. Instead, the rational parts must contain poles. Such Lorentz structures consistent with the overall dimension 6 scaling of the rational part are, up to conjugates: a) two 3-point dimension 5 vertices (\vp \vp \s, \vp \fp \fp, and conjugates) connected by a single propagator, specifically:
\begin{align}
\vp \vp (\s + \s) \vm \vm &= \vp \vp \vm \vm &\text{ with pole in }&  (p[\vp_1 ]+p[\vp_2 ])^2, \\
\vp \fp (\fp + \fm) \fm \vm &= \vp \vm \fp \fm &\text{ with pole in }&  (p[\vp ]+p[\fp ])^2, \label{eq:rightpole} \\
\s \vp (\vp + \vm) \vm \s &= \vp \vm \s \s &\text{ with pole in }&  (p[\vp ]+p[\s ])^2, \\
\s \vp (\vp + \vm) \fm \fm &= \vp \fm \fm \s &\text{ with pole in }&  (p[\vp ]+p[\s ])^2, 
\end{align}
or, for non-Abelian sectors only of a theory, b) a 3-point dimension 6 vertex (\vp \vp \vp, and conjugate) connected with a dimension 4 vertex by a single propagator, specifically:
\begin{align}
\vp \vp (\vp + \vm) \vp \vm &= \vp \vp \vp \vm &\text{ with pole in }&  (p[\vp_1 ]+p[\vp_2 ])^2, \\
\vp \vp (\vp + \vm) \fp \fm &= \vp \vp \fp \fm &\text{ with pole in }&  (p[\vp_1 ]+p[\vp_2 ])^2.
\end{align}

At the same time, all of the 4-point graphs that a 4-field operator will contribute to (\ref{eq:1loopamp}) depend on a single kinematic invariant, and it is only this kinematic invariant that will appear in poles in the rational terms $R$. For example, the $\phi^2 \bar \psi \psi D$ operator could contribute to $\vp \vm \fp \fm$ at one loop via
\begin{equation}
\fp \fm (\s \s + \s \s) \vp \vm = \vp \vm \fp \fm ,
\end{equation}
but it could not generate a valid rational part, as it could only have the (wrong) pole in $(p[\fp]+p[\fm])^2$, rather than the $(p[\vp]+p[\fp])^2$ in (\ref{eq:rightpole}). This logic can account for all but three of the zero rational parts (where there naively exists a diagram) in Table \ref{tab:rationalResults}. The remaining three zeroes, labelled with asterisks, could in principle generate rational parts with the correct poles, namely
\begin{align}
\fm \fm ( \fp \fp + \fm \fm ) \vp \s = \vp \fm \fm \s  
\end{align}
whose graphs are functions of $(p[\vp]+p[\s])^2$, and, in the non-Abelian case,
\begin{align}
\vp \vp ( \s \s + \s \s ) \vp \vm = \vp \vp \vp \vm \, , \\
\vp \vp ( \s \s + \s \s ) \fp \fm = \vp \vp \fp \fm
\end{align}
whose graphs are both functions of $(p[\vp_1]+p[\vp_2])^2$. We do not provide a single argument to explain the starred zeroes in Table \ref{tab:4pointtree} in these three cases.

In short, we find --- perhaps surprisingly --- that the rational part of dimension 6 loops vanishes in many cases, implying that the non-interference theorems are particularly robust against some radiative corrections. All one-loop contributions to the $|h| = 0$ amplitudes in (\ref{4noninterf}) involving an insertion of a $(4,0)$ dimension 6 operator vanish \emph{exactly}. More broadly, the appearance of so many vanishing one-loop dimension 6 helicity amplitudes suggests that the surprising one-loop structure of EFTs at dimension 6 extends beyond logarithms to rational parts and warrants further investigation. We note that the absence of rational parts would also have further implications for the higher-loop structure of the anomalous dimension matrix, which have already been shown to contain many zeroes \cite{Bern:2019wie}.



\section{Phenomenological implications\label{sec:convolve}}

Let us now briefly examine the the phenomenological implications from comparing and convolving the tree/loop classification of operators in \S\ref{sec:treeloop} and the pattern of tree/loop amplitudes in \S\ref{sec:treeloopamp}.

First, consider the pattern of red tree-level generated operators in Figure \ref{fig:dim6treeloop}, which mimicks the conical shape and pattern of the tree-level helicity amplitudes at dimension 4, shown schematically in Figure \ref{fig:dim4map}. In fact, all contact interactions of tree-level generated dimension 6 operators interfere with the corresponding dimension 4 process at tree level; all contact interactions of the loop-level operators do not. Note this correspondence includes the exceptional amplitude $(\psi^+)^4$, which interferes with the tree-level operator $\psi^4$.


The same approximately conical shape means that tree-level operators only renormalise other tree-level operators, with one exception \cite{Elias-Miro:2014eia} --- $\psi^4$ renormalises $F \psi^2 \phi$ , which is highlighted in blue in the Figure. The phenomenological implications of this were pointed out in \cite{Elias-Miro:2014eia} for the case of corrections to the $h \to \gamma\gamma$ rate in the Standard Model \cite{Elias-Miro:2013gya}. Corrections to $h \to \gamma\gamma$ are mediated by an operator of class $\phi^2 F^2$, a loop-level suppressed operator which does not receive, through running, a large log enhanced correction from tree-level operators which could spoil its suppression.

From Figure \ref{fig:dim8treeloop}, we see that qualified versions of the above statements also hold at dimension 8. The set of tree-level operators retain their approximate conical shape. All tree-level generated dimension 8 operators interfere with dimension 4 processes; many, but not all, loop-level operators do not. Tree-level operators mostly renormalise tree-level operators, with a few more blue exceptions: $F \bar F \phi^2 D^2$, $\bar F \psi^2 \phi D^2$, $F \bar F \psi \bar \psi D$, $F \bar \psi^2 \phi D^2$, $F \psi^2 \phi D^2$, $F^2 \bar \psi^2 \phi$, $F^2 \psi^2 \phi$.

Amongst the many dimension 8 operators, one may find phenomenologically relevant analogues of the dimension 6 effects in $h \to \gamma\gamma$. The operators that contribute at leading order to anomalous neutral triple gauge boson couplings ($ZZ\gamma$ and $Z\gamma\gamma$ processes) are of the form $F^2 \phi^2 D^2$ \cite{Degrande:2013kka}. These aNTGC operators are loop-level in weakly coupled completions, and by helicity arguments can only be renormalised by operators with the same coordinates $(4,\pm 2)$. As there does not exist a diagram for the tree-level operators of the class $\psi^4 D^2$ to renormalise the $F^2 \phi^2 D^2$ operators, the loop-level aNTGC operators are only renormalised by loop-level operators, again avoiding the potentially large logarithmic enhancement that would come from the running of a tree-level operator.


Consider now the pattern of tree- and loop-level dimension 6 operators' rational one-loop contributions to the processes (\ref{4noninterf}) and (\ref{42noninterf}). Dimension 6 processes do not contribute to (\ref{4noninterf}) and, in the Abelian case, to (\ref{42noninterf}) at tree-level in the high energy limit. In \cite{Azatov:2016sqh} it was noted that finite mass effects typically provide the leading violation of this non-interference, exceeding possible radiative violations. However, this need not be the case in EFTs with perturbative UV completions whose Wilson coefficients can be classified as tree- and loop-generated. Violation of the non-interference theorem due to finite mass effects arises from applying mass-suppressed helicity flips to tree-level diagrams involving a dimension 6 operator. For the $|h| = 0$ helicity amplitudes in (\ref{4noninterf}), the relevant dimension 6 operators are all loop-generated. Thus the interference terms are suppressed by both $\sim m^2/E^2$ (coming from mass insertions) and a loop factor (coming from the loop suppression of the Wilson coefficient) despite involving only tree-level amplitudes in the EFT.

If there were instead nonzero contributions to the $|h| = 0$ amplitudes in (\ref{4noninterf}) from one loop diagrams containing insertions of tree-level dimension 6 operators, these would comprise the {\it leading} violation of the non-interference theorems, exceeding the contributions due to finite mass effects by a relative factor of $\sim E^2/m^2$, the loop factors being comparable. While such diagrams are necessarily rational at one loop, one might expect them to be nonzero in analogy with their dimension 4 counterparts. Surprisingly, as seen from Table \ref{tab:rationalResults}, this is not the case, and the non-interference is in this sense more robust than expected in the face of radiative corrections.


\section{Discussion\label{sec:conc}}

We have presented a simple method for enumerating the classes of Lorentz structures of operators that arise in a generic massless EFT of scalars, fermions and vectors. Working up to dimension 8, we determined the tree/loop structure of both the operator coefficients (appropriate for perturbative UV completions) and the helicity amplitudes they induce within the EFT, in the process both extending various known dimension 6 results to dimension 8 and discovering new structure in rational amplitudes at dimension 6. 

Functional techniques significantly simplify the tree/loop classification of operator coefficients in perturbative UV completions. Here we have integrated out at tree level a generic weakly coupled renormalisable UV theory of scalars, fermions and vectors (such as arises from a spontaneously broken gauge theory), and considered the resulting pattern of tree- and loop-level generated operators at dimension 6 and 8 within the EFT in the space of $(n, \sum h)$. The arrangement of tree-level operators mirrors that of the dimension 4 tree-level processes, and allows us to rederive known results at dimension 6, and see that they also hold substantively at dimension 8. To wit, it is difficult to generate operators containing field strengths at tree level; tree-level generated operators interfere with dimension 4 tree-level processes whereas loop-level generated operators generally do not; tree-level operators tend to renormalise tree-level operators. One may use these observations to quickly identify the relevant operators contributing significantly to a particular process at one-loop, as we illustrated for the case of anomalous neutral triple gauge boson scattering within the Standard Model.

The tree/loop structure of helicity amplitudes is equally rich. The dimension 4 tree-level $n \geq 4$ processes are cone shaped: other than the exceptional amplitudes arising from the all-plus or all-minus four fermion process, they all lie within a cone $\abs{\sum h} = n -4$, with apex $(4,0)$. By simple construction rules (Figures \ref{fig:treeRule} and \ref{fig:looprule}) \cite{Cheung:2015aba}, one can see that the tree-level contributions of a dimension $d$ operator to dimension $d$ processes therefore lie in a cone with an apex given by the $(n, \sum h)$ coordinates of the operator. The cut constructible one-loop effects of the dimension $d$ operator --- ignoring exceptional amplitudes --- lie within the same cone; the effects of the operator outside of the cone are conversely purely rational (i.e.\ the process has no possible logarithmic enhancement at one-loop).

In general, a one-loop helicity amplitude may have a rational part, even if the corresponding tree-level process vanishes on-shell. However, by explicit computation we have shown this not to be the case for many important processes at dimension 6. Tree-level generated $n=4$ operators have a vanishing one-loop contribution to the processes which vanish on-shell at tree-level. This observation may be of phenomenological use when working to one loop order within the Standard Model EFT: in a weakly coupled UV completion, such dimension 6 processes effectively vanish completely at one loop, notwithstanding mass effects. More broadly, the vanishing of various one loop helicity amplitudes at dimension 6 suggests that the effects of ostensibly tree-level helicity selection rules extends beyond one loop, influencing the two-loop matrix of anomalous dimensions.

Many of these structures lack a more holistic explanation. It is possible that this could be afforded by new techniques to generate bases of on-shell amplitudes, wherein ostensibly different Lorentz structures can be viewed as different elements of representations of an enlarged little group symmetry \cite{Henning:2019mcv,Henning:2019enq}. Furthermore, there are many coordinates that may be associated with a given helicity amplitude (such as the total helicity of external vectors and fermions) which combine additively when constructing amplitudes on-shell. In this sense, the space $(n,\sum h)$ is just a projection within a vector space of much larger dimension, which may conceal additional patterns and clues.

\acknowledgments

We thank S.~Koren, T.~Melia, T.~Trott and Z.~Zhang for helpful discussions. The work of NC and DS is supported in part by the US Department of Energy under the Early Career Award DE-SC0014129. MJ is supported by the National Natural Science Foundation of China (NSFC) under grant No.11761141011. This project has received funding from the European Union's Horizon 2020 research and innovation programme under the Marie Sk\l{}odowska-Curie grant agreement No.\ 754496. This manuscript has been authored by Fermi Research Alliance, LLC, under Contract No. DE-AC02-07CH11359 with the U.S. Department of Energy, Office of Science, Office of High Energy Physics. YYL thanks the Kavli Institute for Theoretical Physics for a graduate fellowship during the inception of this work, supported in part by the National Science Foundation under Grant No. NSF PHY-1748958. DS thanks SISSA for hospitality.

\bibliographystyle{JHEP}
\bibliography{reference}

\appendix

\section{Simplified models for one loop calculations\label{app:simpmodels}}

In order to illustrate the one-loop structure of dimension 6 helicity amplitudes we construct four simplified models, two Abelian and two non-Abelian, using Dirac fermions. Dimension 6 operators are written in linear combinations that excite particles of definite helicity. Not all dimension 6 operators are included, only those that may induce different Lorentz or colour structures in the amplitudes we consider. An external fermion $\psi^\pm$ can be of type 1 or 2; $\phi$ can stand for a scalar particle or its conjugate. The simplified models are: 

1) The Abelian gauge theory of a unit charge fermion $\psi_1$ and scalar $\phi$, along with a neutral $\psi_2$, with
\begin{align}
\lag =& - \frac14 B^2 + i \bar \psi_1 \slashed{D} \psi_1 + i \bar \psi_2 \slashed{D} \psi_2 + \abs{ D \phi}^2 - y \bar \psi_1 \psi_2 \phi + \text{h.c.} \nonumber \\
&+ c_R \bar \phi \phi \abs{ D \phi}^2
+ c_{FF}^+ \bar \phi \phi B (B-\tilde B)
+ c_{FF}^- \bar \phi \phi B (B+\tilde B) \nonumber \\
&+ c_{PP1} i \bar \phi \overset{\leftrightarrow}{D_\mu} \phi \bar \psi_1 \gamma^\mu \psi_1
+ c_{PP2} i \bar \phi \overset{\leftrightarrow}{D_\mu} \phi \bar \psi_2 \gamma^\mu \psi_2  \nonumber \\
& + c_D^+ \bar \psi_1 \sigma^{\mu\nu} P_L \psi_2 \phi B_{\mu\nu} 
+ c_D^- \bar \psi_1 \sigma^{\mu\nu} P_R \psi_2 \phi B_{\mu\nu} \nonumber \\
&+ c_{11LL} \bar \psi_1 \gamma^\mu P_L \psi_1 \bar \psi_1 \gamma^\mu P_L \psi_1
+ c_{12LL} \bar \psi_1 \gamma^\mu P_L \psi_1 \bar \psi_2 \gamma^\mu P_L \psi_2
+ c_{22LL} \bar \psi_2 \gamma^\mu P_L \psi_2 \bar \psi_2 \gamma^\mu P_L \psi_2  \nonumber \\
&+ c_{12LR} \bar \psi_1 \gamma^\mu P_L \psi_1 \bar \psi_2 \gamma^\mu P_R \psi_2
+ c_{12RL} \bar \psi_1 \gamma^\mu P_R \psi_1 \bar \psi_2 \gamma^\mu P_L \psi_2  \nonumber \\
&+ c_{11S}^+ \bar \psi_1 P_L \psi_1 \bar \psi_1 P_L \psi_1 
+ c_{11S}^- \bar \psi_1 P_R \psi_1 \bar \psi_1 P_R \psi_1 
\end{align}
resulting in the following contributions to one-loop dimension 6 helicity amplitudes:\newline
\begin{tabular}{c c c c c c c | c c}
 & & & \multicolumn{4}{c|}{$(4,0)$} & \multicolumn{2}{c}{$(4,2)$}  \\
 & & & \rotp{$V^+ V^+ V^- V^-$} 
 & \rotp{$V^+ V^- \psi^+ \psi^-$}
 & \rotp{$V^+ V^- \phi \phi$}
 & \rotp{$V^+ \psi^- \psi^- \phi$}
 & \rotp{$V^+ V^+ V^+ V^-$}
 & \rotp{$V^+ V^+ \psi^+ \psi^- $}
 \\
\multirow{8}*{$(4,0)$} &
$\phi^4 D^2$ & $c_R$ & \pnd & \pnd & \pzero & \pnd 
 & \pnd & \pnd \\
& \multirow{2}*{$\phi^2 \bar \psi \psi D$} &
$c_{PP1}$ & \pnd & \pzero & \pzero & \pzero 
 & \pnd & \pzero \\
& & $c_{PP2}$ & \pnd & \pzero & \pzero & \pzero 
 & \pnd & \pzero \\
& \multirow{5}*{$\bar \psi^2 \psi^2$} &
 $c_{11LL}$ & \pnd & \pzero & \pnd & \pnd
 & \pnd & \pzero \\
& & $c_{12LL}$ & \pnd & \pzero & \pnd & \pzero
 & \pnd & \pzero \\
& & $c_{22LL}$ & \pnd & \pnd & \pnd & \pnd
 & \pnd & \pnd \\
& & $c_{12LR}$ & \pnd & \pzero & \pnd & \pzero
 & \pnd & \pzero \\
& & $c_{12RL}$ & \pnd & \pzero & \pnd & \pzero
 & \pnd & \pzero \\
 \hline
\multirow{3}*{$(4,2)$} &
 $F \psi^2 \phi$ & $c_D^+$ & \pnd & \pnz & \pnz & \pzero
 & \pnd & \pzero \\
& $F^2 \phi^2$ & $c_{FF}^+$ & \pnz & \pzero & \pnz & \pnz 
 & \pzero & \pzero \\
& $\psi^4$ & $c_{11S}^+$ & \pnd & \pzero & \pnd & \pnd
 & \pnd & \pzero \\
 \hline
\multirow{3}*{$(4,-2)$} &
 $\bar F \bar \psi^2 \phi$ & $c_D^-$ & \pnd & \pnz & \pnz & \pnz
 & \pnd & \pzero \\
& $\bar F^2 \phi^2$ & $c_{FF}^-$ & \pnz & \pzero & \pnz & \pzero 
 & \pzero & \pzero \\
& $\bar \psi^4$ & $c_{11S}^-$ & \pnd & \pzero & \pnd & \pnd
 & \pnd & \pzero \\
\end{tabular}

The key for the above table is as in Table \ref{tab:rationalResults}. \\

2) The Abelian gauge theory of a neutral real scalar, and two unit charge fermions $\psi_1$ and $\psi_2$, with
\begin{align}
\lag =& - \frac14 B^2 + i \bar \psi_1 \slashed{D} \psi_1 + i \bar \psi_2 \slashed{D} \psi_2 + \frac12 ( \partial \phi)^2 - y \bar \psi_1 \psi_2 \phi + \text{h.c.} \nonumber \\
&+ c_{FF}^+ \bar \phi \phi B (B-\tilde B)
+ c_{FF}^- \bar \phi \phi B (B+\tilde B) \nonumber \\
&+ c_D^+ \bar \psi_1 \sigma^{\mu\nu} P_L \psi_2 \phi B_{\mu\nu} 
+ c_D^- \bar \psi_1 \sigma^{\mu\nu} P_R \psi_2 \phi B_{\mu\nu} \nonumber \\
&+ c_{11LL} \bar \psi_1 \gamma^\mu P_L \psi_1 \bar \psi_1 \gamma^\mu P_L \psi_1
+ c_{12LL} \bar \psi_1 \gamma^\mu P_L \psi_1 \bar \psi_2 \gamma^\mu P_L \psi_2
+ c_{12LR} \bar \psi_1 \gamma^\mu P_L \psi_1 \bar \psi_2 \gamma^\mu P_R \psi_2 \nonumber \\
&+ c_{12LL\times} \bar \psi_2 \gamma^\mu P_L \psi_1 \bar \psi_2 \gamma^\mu P_L \psi_1 + \text{h.c.}  
+ c_{12S}^+ \bar \psi_2 P_L \psi_2 \bar \psi_1 P_L \psi_1 
+ c_{12S}^- \bar \psi_2 P_R \psi_2 \bar \psi_1 P_R \psi_1 
\end{align}
resulting in \newline
\begin{tabular}{c c c c c c c | c c}
 & & & \multicolumn{4}{c|}{$(4,0)$} & \multicolumn{2}{c}{$(4,2)$}  \\
 & & & \rotp{$V^+ V^+ V^- V^-$} 
 & \rotp{$V^+ V^- \psi^+ \psi^-$}
 & \rotp{$V^+ V^- \phi \phi$}
 & \rotp{$V^+ \psi^- \psi^- \phi$}
 & \rotp{$V^+ V^+ V^+ V^-$}
 & \rotp{$V^+ V^+ \psi^+ \psi^- $}
 \\
\multirow{4}*{$(4,0)$} &
 \multirow{4}*{$\bar \psi^2 \psi^2$} &
 $c_{11LL}$ & \pnd & \pzero & \pnd & \pnd
 & \pnd & \pzero \\
& & $c_{12LL}$ & \pnd & \pzero & \pnd & \pzero
 & \pnd & \pzero \\
& & $c_{12LL\times}$ & \pnd & \pnd & \pnd & \pzero
 & \pnd & \pnd \\
& & $c_{12LR}$ & \pnd & \pzero & \pnd & \pzero
 & \pnd & \pzero \\
 \hline
\multirow{3}*{$(4,2)$} &
 $F \psi^2 \phi$ & $c_D^+$ & \pnd & \pnz & \pnz & \pnz
 & \pnd & \pzero \\
& $F^2 \phi^2$ & $c_{FF}^+$ & \pnd & \pzero & \pzero & \pnz 
 & \pzero & \pzero \\
& $\psi^4$ & $c_{12S}^+$ & \pnd & \pzero & \pnd & \pzero
 & \pnd & \pzero \\
 \hline
\multirow{3}*{$(4,-2)$} &
 $\bar F \bar \psi^2 \phi$ & $c_D^-$ & \pnd & \pnz & \pnz & \pzero
 & \pnd & \pzero \\
& $\bar F^2 \phi^2$ & $c_{FF}^-$ & \pnd & \pzero & \pzero & \pzero 
 & \pzero & \pzero \\
& $\bar \psi^4$ & $c_{12S}^-$ & \pnd & \pzero & \pnd & \pnz
 & \pnd & \pzero \\
\end{tabular} \\

3) The $SU(3)$ gauge theory of a colour triplet fermion $\psi_1$ and scalar $\phi$, along with a colour singlet $\psi_2$, with
\begin{align}
\lag =& - \frac14 G^2 + i \bar \psi_1 \slashed{D} \psi_1 + i \bar \psi_2 \slashed{D} \psi_2 + \abs{ D \phi}^2 - y \bar \psi_1 \psi_2 \phi + \text{h.c.} \nonumber \\
&+ c_R \bar \phi \phi \abs{ D \phi}^2
+ c_{R,\Box} \bar \phi \phi \Box (\bar \phi \phi)
+ c_{FF}^+ \bar \phi \phi G (G-\tilde G)
+ c_{FF}^- \bar \phi \phi G (G+\tilde G)  \nonumber \\
&+ c_{PP1} i \bar \phi \overset{\leftrightarrow}{D_\mu} \phi \bar \psi_1 \gamma^\mu \psi_1
+ c_{PP2} i \bar \phi \overset{\leftrightarrow}{D_\mu} \phi \bar \psi_2 \gamma^\mu \psi_2  
+ c_{PP3} i \bar \phi \overset{\leftrightarrow}{D_\mu} T^a \phi \bar \psi_1 \gamma^\mu T^a \psi_1  \nonumber \\
&+ c_D^+ \bar \psi_1 \sigma^{\mu\nu} P_L T^a \psi_2 \phi G^a_{\mu\nu} + \text{h.c.}  
+ c_D^- \bar \psi_1 \sigma^{\mu\nu} P_R T^a \psi_2 \phi G^a_{\mu\nu} + \text{h.c.}  \nonumber \\
&+ c_{11LL} \bar \psi_1 \gamma^\mu P_L \psi_1 \bar \psi_1 \gamma^\mu P_L \psi_1
+ c_{12LL} \bar \psi_1 \gamma^\mu P_L \psi_1 \bar \psi_2 \gamma^\mu P_L \psi_2
+ c_{22LL} \bar \psi_2 \gamma^\mu P_L \psi_2 \bar \psi_2 \gamma^\mu P_L \psi_2  \nonumber \\
&+ c_{11LR(8)} \bar \psi_1 \gamma^\mu P_L T^a \psi_1 \bar \psi_1 \gamma^\mu P_R T^a \psi_1  \nonumber \\
&+ c_{12LR} \bar \psi_1 \gamma^\mu P_L \psi_1 \bar \psi_2 \gamma^\mu P_R \psi_2
+ c_{12RL} \bar \psi_1 \gamma^\mu P_R \psi_1 \bar \psi_2 \gamma^\mu P_L \psi_2 \nonumber \\ 
&+ c_{11S}^+ \bar \psi_1 P_L \psi_1 \bar \psi_1 P_L \psi_1
+ c_{11S}^- \bar \psi_1 P_R \psi_1 \bar \psi_1 P_R \psi_1
\end{align}
resulting in \newline
\begin{tabular}{c c c c c c c | c c}
 & & & \multicolumn{4}{c|}{$(4,0)$} & \multicolumn{2}{c}{$(4,2)$}  \\
 & & & \rotp{$V^+ V^+ V^- V^-$} 
 & \rotp{$V^+ V^- \psi^+ \psi^-$}
 & \rotp{$V^+ V^- \phi \phi$}
 & \rotp{$V^+ \psi^- \psi^- \phi$}
 & \rotp{$V^+ V^+ V^+ V^-$}
 & \rotp{$V^+ V^+ \psi^+ \psi^- $}
 \\
\multirow{10}*{$(4,0)$} &
\multirow{2}*{$\phi^4 D^2$} &
$c_R$ & \pnd & \pnd & \pzero & \pzero 
 & \pnd & \pnd \\
& & $c_{R,\Box}$ & \pnd & \pnd & \pzero & \pzero 
 & \pnd & \pnd \\
& \multirow{3}*{$\phi^2 \bar \psi \psi D$} &
$c_{PP1}$ & \pnd & \pzero & \pzero & \pzero 
 & \pnd & \pzero \\
& & $c_{PP2}$ & \pnd & \pzero & \pzero & \pzero 
 & \pnd & \pzero \\
& & $c_{PP3}$ & \pnd & \pzero & \pzero & \pzero 
 & \pnd & \pnz \\
& \multirow{6}*{$\bar \psi^2 \psi^2$} &
 $c_{11LL}$ & \pnd & \pzero & \pnd & \pnd
 & \pnd & \pzero \\
& & $c_{12LL}$ & \pnd & \pzero & \pnd & \pzero
 & \pnd & \pzero \\
& & $c_{22LL}$ & \pnd & \pnd & \pnd & \pnd
 & \pnd & \pnd \\
& & $c_{12LR}$ & \pnd & \pzero & \pnd & \pzero
 & \pnd & \pzero \\
& & $c_{12RL}$ & \pnd & \pzero & \pnd & \pzero
 & \pnd & \pzero \\
& & $c_{11LR(8)}$ & \pnd & \pzero & \pnd & \pnd
 & \pnd & \pnz \\
 \hline
\multirow{3}*{$(4,2)$} &
 $F \psi^2 \phi$ &
$c_D^+$ & \pnd & \pnz & \pnz & \pzero
 & \pnd & \pzero \\
& $F^2 \phi^2$ & $c_{FF}^+$ 
& \pnz & \pzero & \pnz & \pnz 
 & \pzero & \pzero \\
& $\psi^4$ & $c_{11S}^+$ & \pnd & \pzero & \pnd & \pnd
 & \pnd & \pzero \\
 \hline
\multirow{4}*{$(4,-2)$} &
 $\bar F \bar \psi^2 \phi$ &
$c_D^-$ & \pnd & \pnz & \pnz & \pnz
 & \pnd & \pzero \\
& $\bar F^2 \phi^2$ & $c_{FF}^-$ 
& \pnz & \pzero & \pnz & \pzero 
 & \pzero & \pzero \\ 
& $\bar \psi^4$ & $c_{11S}^-$ & \pnd & \pzero & \pnd & \pnd
 & \pnd & \pzero \\
\end{tabular}

4) The $SU(3)$ gauge theory of a singlet real scalar, and colour triplet fermions $\psi_1$ and $\psi_2$, with
\begin{align}
\lag =& - \frac14 G^2 + i \bar \psi_1 \slashed{D} \psi_1 + i \bar \psi_2 \slashed{D} \psi_2 + \frac12 (\partial \phi)^2 - y \bar \psi_1 \psi_2 \phi + \text{h.c.} \nonumber \\
&+ c_{FF}^+ \phi \phi G (G-\tilde G)
+ c_{FF}^- \phi \phi G (G+\tilde G)  \nonumber \\
&+ c_D^+ \bar \psi_1 \sigma^{\mu\nu} P_L T^a \psi_2 \phi G^a_{\mu\nu} + \text{h.c.} 
+ c_D^- \bar \psi_1 \sigma^{\mu\nu} P_R T^a \psi_2 \phi G^a_{\mu\nu} + \text{h.c.}  \nonumber \\
&+ c_{11LL} \bar \psi_1 \gamma^\mu P_L \psi_1 \bar \psi_1 \gamma^\mu P_L \psi_1
+ c_{12LL} \bar \psi_1 \gamma^\mu P_L \psi_1 \bar \psi_2 \gamma^\mu P_L \psi_2  \nonumber \\
&+ c_{12LL(8)} \bar \psi_1 \gamma^\mu P_L T^a \psi_1 \bar \psi_2 \gamma^\mu P_L T^a \psi_2
+ c_{12LL\times} \bar \psi_2 \gamma^\mu P_L \psi_1 \bar \psi_2 \gamma^\mu P_L \psi_1 + \text{h.c.}  \nonumber \\
&+ c_{12LR} \bar \psi_1 \gamma^\mu P_L \psi_1 \bar \psi_2 \gamma^\mu P_R \psi_2
+ c_{12LR(8)} \bar \psi_1 \gamma^\mu P_L T^a \psi_1 \bar \psi_2 \gamma^\mu P_R T^a \psi_2 \nonumber \\
&+ c_{12RL} \bar \psi_1 \gamma^\mu P_R \psi_1 \bar \psi_2 \gamma^\mu P_L \psi_2 
+ c_{12RL(8)} \bar \psi_1 \gamma^\mu P_R T^a \psi_1 \bar \psi_2 \gamma^\mu P_L T^a \psi_2 \nonumber \\
&+ c_{12S}^+ \bar \psi_2 P_L \psi_2 \bar \psi_1 P_L \psi_1 
+ c_{12S(8)}^+ \bar \psi_2 P_L T^a \psi_2 \bar \psi_1 P_L T^a \psi_1 \nonumber \\ 
&+ c_{12S}^- \bar \psi_2 P_R \psi_2 \bar \psi_1 P_R \psi_1 
+ c_{12S(8)}^- \bar \psi_2 P_R T^a \psi_2 \bar \psi_1 P_R T^a \psi_1 
\end{align}
resulting in \newline
\begin{tabular}{c c c c c c c | c c}
 & & & \multicolumn{4}{c|}{$(4,0)$} & \multicolumn{2}{c}{$(4,2)$}  \\
 & & & \rotp{$V^+ V^+ V^- V^-$} 
 & \rotp{$V^+ V^- \psi^+ \psi^-$}
 & \rotp{$V^+ V^- \phi \phi$}
 & \rotp{$V^+ \psi^- \psi^- \phi$}
 & \rotp{$V^+ V^+ V^+ V^-$}
 & \rotp{$V^+ V^+ \psi^+ \psi^- $}
 \\
\multirow{8}*{$(4,0)$} 
& \multirow{8}*{$\bar \psi^2 \psi^2$} &
 $c_{11LL}$ & \pnd & \pzero & \pnd & \pnd
 & \pnd & \pzero \\
& & $c_{12LL}$ & \pnd & \pzero & \pnd & \pzero
 & \pnd & \pzero \\
& & $c_{12LR}$ & \pnd & \pzero & \pnd & \pzero
 & \pnd & \pzero \\
& & $c_{12LR(8)}$ & \pnd & \pzero & \pnd & \pzero
 & \pnd & \pnz \\
& & $c_{12RL}$ & \pnd & \pzero & \pnd & \pzero
 & \pnd & \pzero \\
& & $c_{12RL(8)}$ & \pnd & \pzero & \pnd & \pzero
 & \pnd & \pnz \\
& & $c_{12LL(8)}$ & \pnd & \pzero & \pnd & \pzero
 & \pnd & \pzero \\
& & $c_{12LL\times}$ & \pnd & \pnd & \pnd & \pzero
 & \pnd & \pnd \\
 \hline
\multirow{4}*{$(4,2)$} &
 $F \psi^2 \phi$ &
$c_D^+$ & \pnd & \pnz & \pnz & \pnz
 & \pnd & \pzero \\
& $F^2 \phi^2$ & $c_{FF}^+$ 
& \pzero & \pzero & \pzero & \pnz 
 & \pzero & \pzero \\
& \multirow{2}*{$\psi^4$} & $c_{12S}^+$ & \pnd & \pzero & \pnd & \pzero
 & \pnd & \pzero \\
&  & $c_{12S(8)}^+$ & \pnd & \pzero & \pnd & \pzero
 & \pnd & \pzero \\
 \hline
\multirow{4}*{$(4,-2)$} &
 $\bar F \bar \psi^2 \phi$ &
$c_D^-$ & \pnd & \pnz & \pnz & \pzero
 & \pnd & \pzero \\
& $\bar F^2 \phi^2$ & $c_{FF}^-$ 
& \pzero & \pzero & \pzero & \pzero
 & \pzero & \pzero \\ 
& \multirow{2}*{$\bar \psi^4$} & $c_{12S}^-$ & \pnd & \pzero & \pnd & \pnz
 & \pnd & \pzero \\
&  & $c_{12S(8)}^-$ & \pnd & \pzero & \pnd & \pnz
 & \pnd & \pzero \\ 
\end{tabular}

\end{document}